\documentclass[times,twocolumn,final]{elsarticle}

\usepackage{framed,multirow}

\usepackage{amssymb}
\usepackage{latexsym}

\usepackage{url}
\usepackage{xcolor}
\usepackage{amsmath}
\usepackage{array} 
\usepackage{tabularx}
\usepackage{ragged2e}
\usepackage{booktabs}
\usepackage{rotating}
\usepackage{float}
\usepackage{lscape}
\usepackage{pdflscape}
\usepackage{supertabular}
\usepackage{longtable}
\usepackage[a4paper, top=1in, bottom=1.0in, left=0.75in, right=0.75in]{geometry}
\usepackage{changepage}
\usepackage[hidelinks]{hyperref} 
\usepackage{hyperref}
\usepackage{soul}
\usepackage{stfloats}
\hypersetup{
  colorlinks   = true, 
  urlcolor     = blue, 
  linkcolor    = blue, 
  citecolor   = red 
}
\definecolor{newcolor}{rgb}{.8,.349,.1}

\begin{document}


\begin{frontmatter}


\title{From Traditional to Deep Learning Approaches in Whole Slide Image Registration: A Methodological Review}%


\author[1]{Behnaz Elhaminia\corref{cor1}}
\cortext[cor1]{Corresponding author}
\ead{behnaz.elhaminia@warwick.ac.uk}

\author[1]{Abdullah Alsalemi}
\author[1]{Esha Nasir}
\author[1]{Mostafa Jahanifar}
\author[1]{Ruqayya Awan}
\author[2]{Lawrence S. Young}
\author[1,3]{Nasir M. Rajpoot}
\author[1]{Fayyaz Minhas}
\author[1]{Shan E Ahmed Raza}
\ead{shan.raza@warwick.ac.uk}

\address[1]{Tissue Image Analytics (TIA) Centre, Department of Computer Science, University of Warwick, UK.}
\address[2]{Division of Biomedical Sciences, Warwick Medical School, University of Warwick, UK.}
\address[3]{Histofy Ltd, Coventry, UK.}



\begin{abstract}

Whole slide image (WSI) registration is an essential task for analysing the tumour microenvironment (TME) in histopathology. It involves the alignment of spatial information between WSIs of the same section or serial sections of a tissue sample. The tissue sections are usually stained with single or multiple biomarkers before imaging, and the goal is to identify neighbouring nuclei along the Z-axis for creating a 3D image or identifying subclasses of cells in the TME. This task is considerably more challenging compared to radiology image registration, such as magnetic resonance imaging or computed tomography, due to various factors. These include gigapixel size of images, variations in appearance between differently stained tissues, changes in structure and morphology between non-consecutive sections, and the presence of artefacts, tears, and deformations.
Currently, there is a noticeable gap in the literature regarding a review of the current approaches and their limitations, as well as the challenges and opportunities they present. We aim to provide a comprehensive understanding of the available approaches and their application for various purposes.
Furthermore, we investigate current deep learning methods used for WSI registration, emphasising their diverse methodologies. We examine the available datasets and explore tools and software employed in the field. Finally, we identify open challenges and potential future trends in this area of research.

\end{abstract}


\end{frontmatter}

\newcolumntype{P}[1]{>{\RaggedRight\footnotesize}p{#1}}
\section{Introduction}
\label{sec:introduction}

Image registration, also known as image alignment, is a technique to find a spatial transformation for aligning two or more images. This transformation brings the images into a common coordinate system, making them directly comparable. For various tasks, when working with images of the same object or a scene, acquired at different times, from different sensors, or under different conditions, it is necessary to register them on the same coordinate system as they would in a physical space.
In the field of medicine, image analysis plays a pivotal role in diagnosis, prognosis, treatment planning, and follow-up monitoring.
However, in the majority of the cases, the images are acquired using multiple modalities, markers or techniques which vary in terms of temporal, spatial and dimensional aspects which necessitates the need to align the images for appropriate diagnosis and downstream analysis. Consequently, image registration has emerged as an indispensable tool for medical image analysis and has been used in many applications \cite{fu2020deep}.

\begin{figure*}
    \centering
    \includegraphics[width=11.75cm]{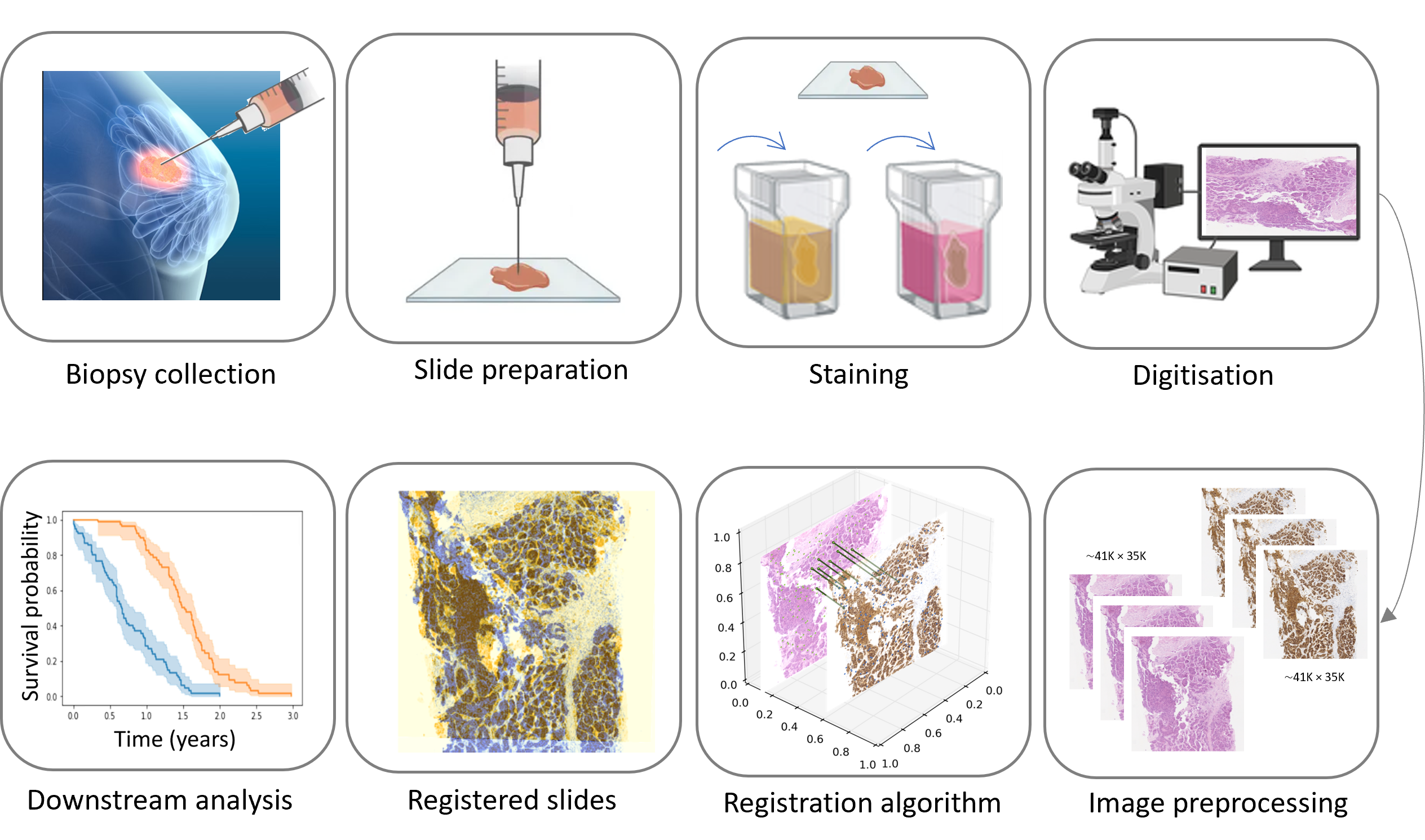}
    \caption{Overview of a general pipeline for WSI registration and downstream analysis. After biopsy collection and digitization, the registration algorithm matches two images. The resulting registered images are then used for further analysis. In this figure, the downstream analysis is adopted from the work of Westein et al. \cite{wetstein2022deep}, which focuses on the analysis of survival rates.}
    \label{fig:reg_general}
\end{figure*}

In pathology, analysing multi-gigapixel images of serial and differently stained histology sections provides valuable insights into the spatial heterogeneity of various molecular markers and the distribution of various kinds of cells. For example, the distribution of tumour cells in relation to immune cell subtypes in the TME can be highly informative for understanding disease dynamics and guiding treatment strategies. Additionally, the quantification of various critical factors, such as the extent of lymphocytic infiltration, aids in understanding disease progression, treatment response, and potential therapeutic strategies. Furthermore, multi-stained tissue slide analysis can help predict responses to immunotherapy using immune markers \cite{wodzinski2021deephistreg}.

Generating a histology WSI includes slicing a tissue sample into thin layers (typically 2-5 $\mu$m thick \cite{lotz2015patch}) and then staining them using various markers to visualise different cellular components. Fig.\ref{fig:reg_general} illustrates the slide preparation. The most common stain is Haematoxylin and Eosin (HE) stain which is commonly used in routine pathology practice \cite{snead2016validation}.

\begin{figure}
    \centering
    \includegraphics[width=7.5cm]{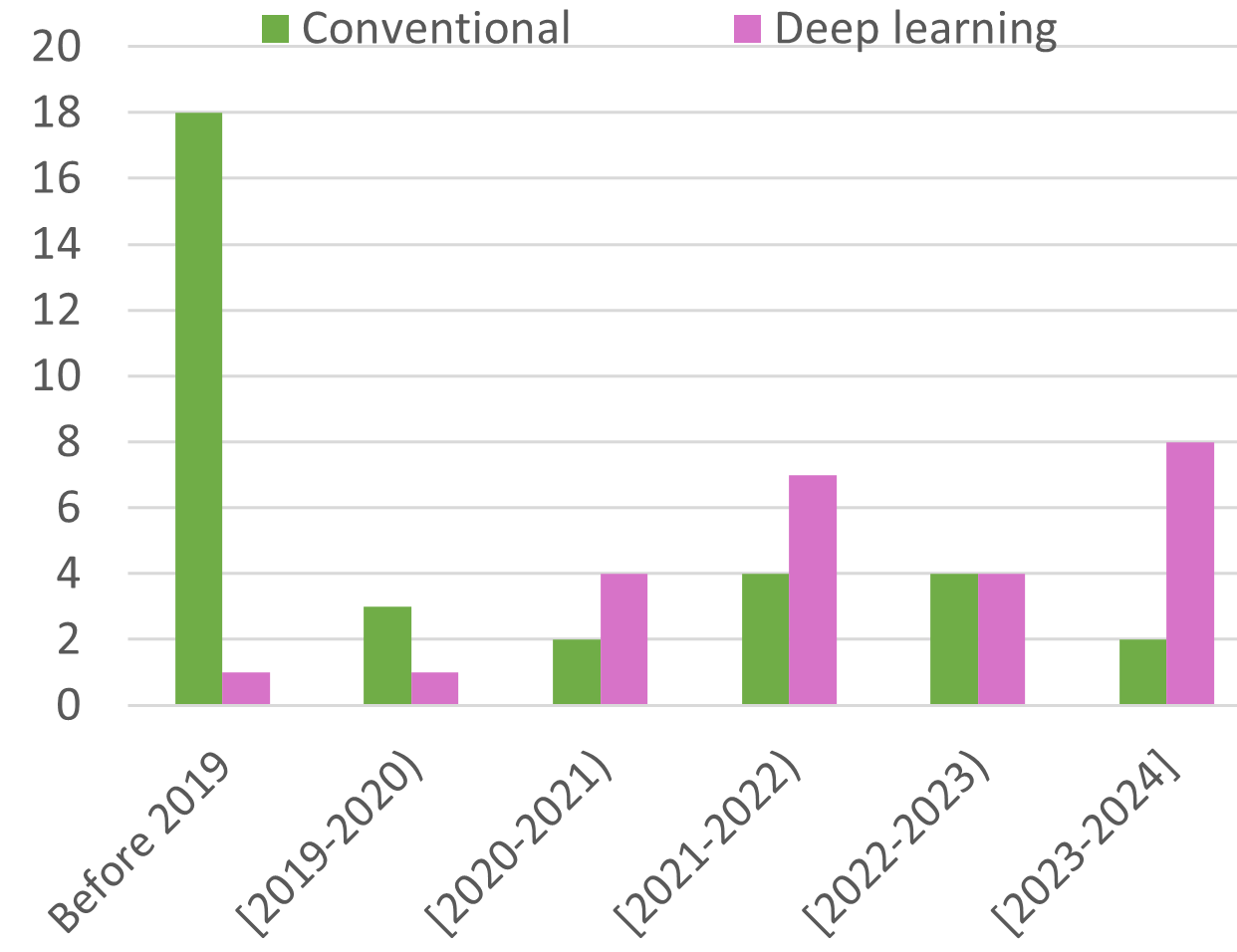}
    \caption{Breakdown of papers proposed for histology image registration (included in this review) in the year of publication. The chart depicts rising trends in the utilisation of deep learning for WSI registration.}
    \label{fig:NumofWorks}
\end{figure}

The stained slides serve as a crucial tool for diverse analyses in pathology. In several tasks, other stains in addition to HE such as immunohistochemical (IHC) stains like Ki67 or Estrogen Receptor (ER), may be required for a comprehensive analysis e.g., for ER scoring in breast cancer \cite{saha2020hscorenet}. Fluorescence imaging further adds strength to these techniques \cite{moustaka2023early}. There are emerging technologies for multiplexed imaging such as CODEX \cite{black2021codex}, which provide a deep view of spatial relationships at the single-cell level within tissues. However, combining information from distinct images such as multiplexed immunofluorescence (mIF) images, serial section images, serially stained images, and other diverse forms of data with routine HE WSIs requires precise multimodal image registration. To meet this demand, numerous researchers have proposed various registration methods, leading to a growing number of papers focusing on this aspect. On the other hand, deep learning has gained considerable attention in recent years for medical image registration \cite{chen2023image}, consequently prompting researchers to employ it for WSI registration. Fig.\ref{fig:NumofWorks} summarises the breakdown of papers published in the area of WSI registration over the past years and included in this review. The chart shows a rise in the utilisation of deep learning for WSI registration.

While numerous surveys have exhaustively covered medical image registration \cite{boveiri2020medical,haskins2020deep,fu2020deep,xiao2021review}, there is a notable gap in the domain of WSI registration. 
Since digital pathology has become more prevalent, it is crucial to address this research gap, and hence a detailed review of state-of-the-art approaches along with the challenges and opportunities presented by WSI registration, needs to be documented.

\begin{figure*}
    \centering
    \includegraphics[width=1.0\textwidth]{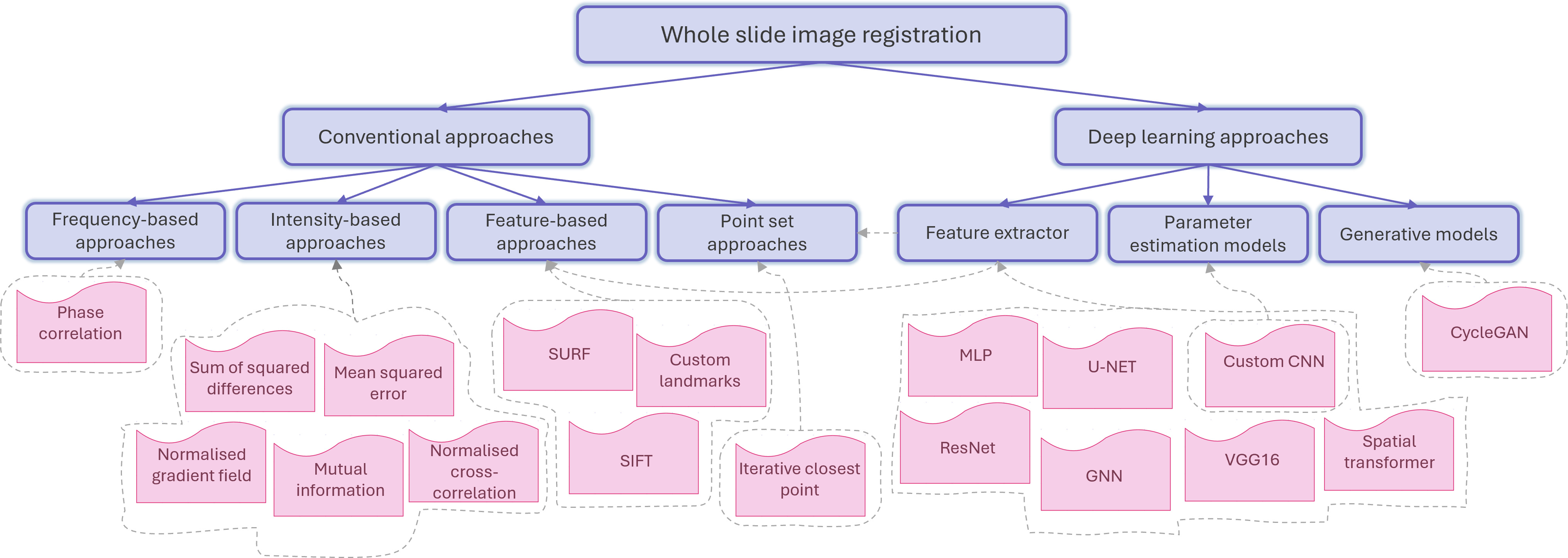}
    \caption{Taxonomy of registration methods reviewed in this study. The methods are categorised based on their approach into two main groups: deep learning models and conventional approaches. General models used for each category and reviewed in this work are depicted in pink.}
    \label{fig:techniques}
\end{figure*}

In this work, we aim to review state-of-the-art research on WSI registration, with a focus on methods using deep learning. 
We categorise the methods into two distinct approaches: conventional methods and deep learning models. Fig.\ref{fig:techniques} shows the classification of the methods along with their common subcategory models for each.
Reviewed literature is investigated from various perspectives such as techniques, evaluation metrics, datasets, and modalities. Deep learning-based methods are grouped into three categories based on their network approaches. Finally, we explore the available datasets, software, and tools, and discuss various challenges, open problems, and possible future directions. This review provides a thorough understanding for readers in the field who are exploring the latest advancements and aiming to contribute to future research.

\section{WSI Registration}
\subsection{Problem Formulation \label{sec:WSIRegistration}}
WSI registration is a process that involves aligning and merging multiple WSI scans obtained from different slides, scanners, or time points. In a conventional setting, the pathologists perform WSI registration manually or semi-automatically. The process involves visually aligning multiple WSIs using specialised software or viewer tools. This requires expert execution and supervision, and it is a time-consuming task that demands meticulous attention to detail and expertise in identifying reference points for alignment. However, with advances in digital pathology technology, automated registration algorithms are increasingly being developed to streamline the process and improve efficiency and accuracy. Fig.\ref{fig:reg_general} illustrates a general pipeline for automatic histopathology image registration and follow-up downstream analysis.

From a mathematical perspective, the standard registration problem can be defined as finding a transformation that optimally aligns an image (moving image) $I_m$ to a reference image $I_r$. With this definition, registration is formulated by defining a cost function $C(I_1, I_2)$ that quantifies the quality of the alignment. Therefore, the objective is to find the transformation $T_{\theta^*}$ that minimises the cost function as:
\begin{equation}
     \theta^* = \arg\min\limits_{\theta} C(I_r, T_\theta(I_m)) ,
\end{equation}
where $\theta$ is the set of transformation parameters.

Despite advancements in automatic registration and the numerous proposed methods by researchers, the registration process remains challenging, whether conducted automatically or manually. Various factors contribute to this complexity, including variations in image quality, differences in imaging modalities, and the presence of noise and artefacts. Additionally, the diverse nature of the objects and scenes being registered further complicates the process, making it a persistent challenge in the field. These challenges and the details of each approach are discussed in the following section.

\subsection{Importance of WSI Registration}
While HE WSIs provide details about morphological characteristics, they are unable to show the expression of different pathologically relevant biomolecules such as proteins for which immunohistochemical markers are necessary. While some techniques allow the characterisation of both morphological and immunohistochemical markers on the same tissue such as the simultaneous staining of PHH3 with HE on the same slide in \cite{jahn2020digital}, typically HE and IHC staining are done on serial sections which warrants effective and reliable registration. Moreover, WSI registration can serve as a means for automated label extraction on a large scale; specific IHC biomarkers, such as PHH3 for mitosis, can be used to stain the same slide that has been stained with HE \cite{ibrahim2023improving}. This allows for quicker or automatic labelling of biological entities in the HE slide based on the biomarker expression in the IHC slide \cite{tellez2018whole, aubreville2024domain}. However, to accomplish this effectively, reliable WSI registration is necessary.

Another significant aspect of WSI registration lies in its application for generating 3D reconstructions from a sequence of 2D slices \cite{kiemen2022coda, arganda2006consistent, feuerstein2011reconstruction, tang2011automatic}. This can help for a more comprehensive understanding of complex structures and spatial relationships within biological specimens.  

Furthermore, merging data from slices stained with varying techniques \cite{obando2017multi, deniz2015multi, trahearn2017hyper, shafique2021automatic}, or different image modalities, such as alignment of histology to magnetic resonance imaging (MRI) \cite{goubran2019multimodal, casamitjana2022robust, shao2021prosregnet}, also emphasise the significance of WSI registration. Generating a high-resolution mosaic from small 2D tiles \cite{pmid20529937},
assisting in virtual staining \cite{bai2023deep} and facilitating for classification of cancer cells \cite{su2022deep} are among other applications of WSI registration.

\subsection{Challenges in WSI Registration}
Generally, finding a transformation that is simultaneously locally and globally consistent in terms of semantics across different images, i.e., similar objects are correctly aligned even in the presence of significant elastic (stretches) or inelastic deformations (tears or missing tissue)  as well as visual inconsistencies such as scanning artefacts, is a challenging task. 

\begin{figure*}[t!]
    \centering
    \includegraphics[width=17cm]{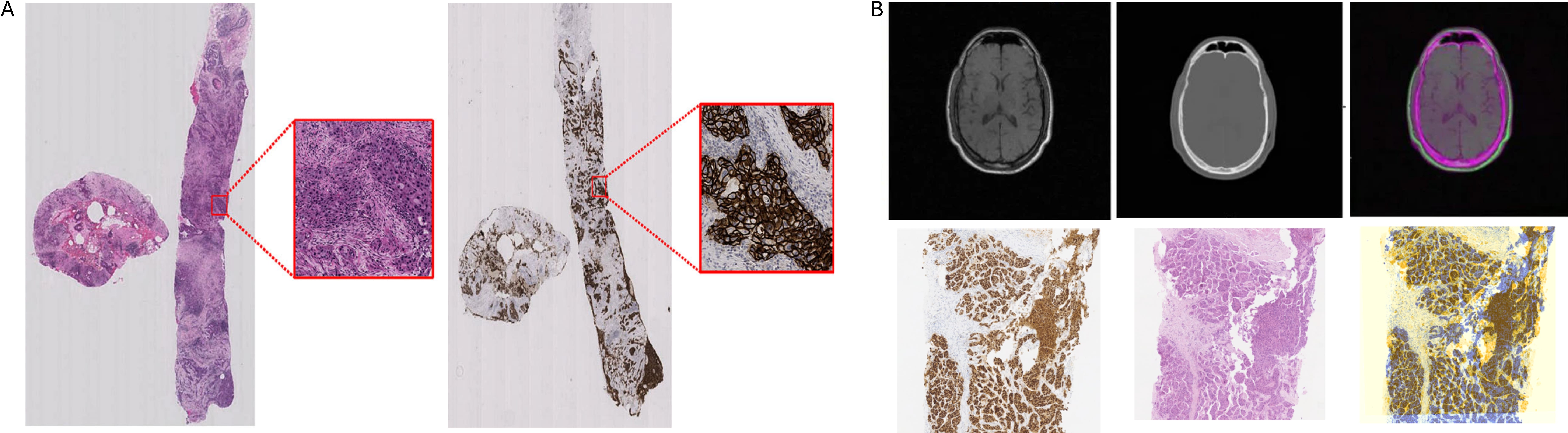}
    \caption{A: An example of WSI with a magnified view of the tumour region. Both views depict the same tissue sample: the left stained with HE for general tissue structure, and the right stained with IHC to highlight specific protein expression \cite{qaiser2018her}. Each image has a resolution of approximately 100K $\times$ 50K pixels. B: Comparison of traditional medical image and WSI registration. The top row displays the registration of CT scan and MRI images. From left to right: CT scan, MRI, and the resulting overlap of registration \cite{gui2023normal}. The bottom row depicts the registration of patches from IHC and HE images. From left to right: IHC image, HE image, and the registration overlapping results displayed as a false-colour image \cite{trahearn2017hyper}\label{fig:mixed}. In radiology image registration, both images often share common baselines that help in alignment, whereas histopathology images typically lack such features, making registration more challenging. }
\end{figure*}

In particular, one of the most challenging parts of WSI registration is to tackle the variability in the image content from the same tissue stained with multiple markers. In most histopathology workflows where WSI registration is required, the slides are marked with stains that highlight different parts of the tissue, e.g., some stains mark the nuclei such as DAPI and Ki67 and others mark the cytoplasm, such as CD163 and CD86 or extracellular matrices such as HE \cite{gurcan2009histopathological}. The batch number of antibodies may further aggravate the variability in the imaging of the stains. This means that prominent features in one stain (e.g., based on nuclei) might be faint or absent in the other, making it difficult to find matching points between the images for alignment, and requiring more sophisticated methods for registration. Fig.\ref{fig:mixed}A shows two different stains applied to a single tissue, revealing significant differences in appearance and the absence of shared patterns.

Additionally, tissues scanned at different time points may exhibit variations in appearance due to changes in tissue structure, intensity, and imaging conditions. Biological processes like growth, deformation, or movement can alter the morphology of tissues.  These differences make it difficult to accurately align images from different time points, as tissues may not retain the same shape, contrast, or position.

Moreover, there may be notable variations in local structures even between serial tissue sections. Variations in section thickness can substantially affect the observed tissue content. This discrepancy poses a significant challenge, as the absence of consistent patterns across slides makes it difficult to find corresponding features between slides \cite{kindle2011semiautomated}. Comparatively, medical images from modalities, such as X-ray(s) and MRI(s) often present inherent features like distinct contours and identifiable structures such as bones, facilitating their registration process. Thus, inherently containing information that aids in aligning them accurately. For instance, Fig.\ref{fig:mixed}B illustrates the comparison between the registration of computed tomography (CT) scans and MRI with WSI. It shows that the contours of the bone are common between MRI and CT. While traditional medical image registration benefits from a shared baseline for bones, aiding in transformation detection, such a common reference point is lacking in WSI due to its varied patterns. Consequently, registration models proposed for traditional medical images often can be applied more universally across various imaging scenarios, while for WSI registration it is challenging; algorithms trained for one specific marker may not extend to other markers. As a result, the development of robust and versatile WSI registration algorithms necessitates careful considerations associated with different staining methods and cellular features \cite{borovec2020anhir}. Given these challenges, registration methods developed for other medical imaging modalities such as X-ray/CT/MRI may not function optimally for WSI registration and require substantial adaptation in addressing the challenges.

Another main challenge is the size of the data; a WSI contains billions of pixels and suffers substantial morphological heterogeneity. The huge size and resolution impose challenges in terms of efficient storage, loading, processing, and analysis; the storage and transmission of entire slide images demand considerable storage capacity and bandwidth. WSIs are primarily stored in gigapixel resolution, offering varying levels of detail; in uncompressed form, a whole slide can exceed the size of 250 Gigabytes, and only high-end machines can handle such amount of data in the system. The computational analysis of these images is therefore a resource-intensive task \cite{gurcan2009histopathological}.

\begin{figure*}[tbh!]
    \centering
    \includegraphics[width=17.0cm]{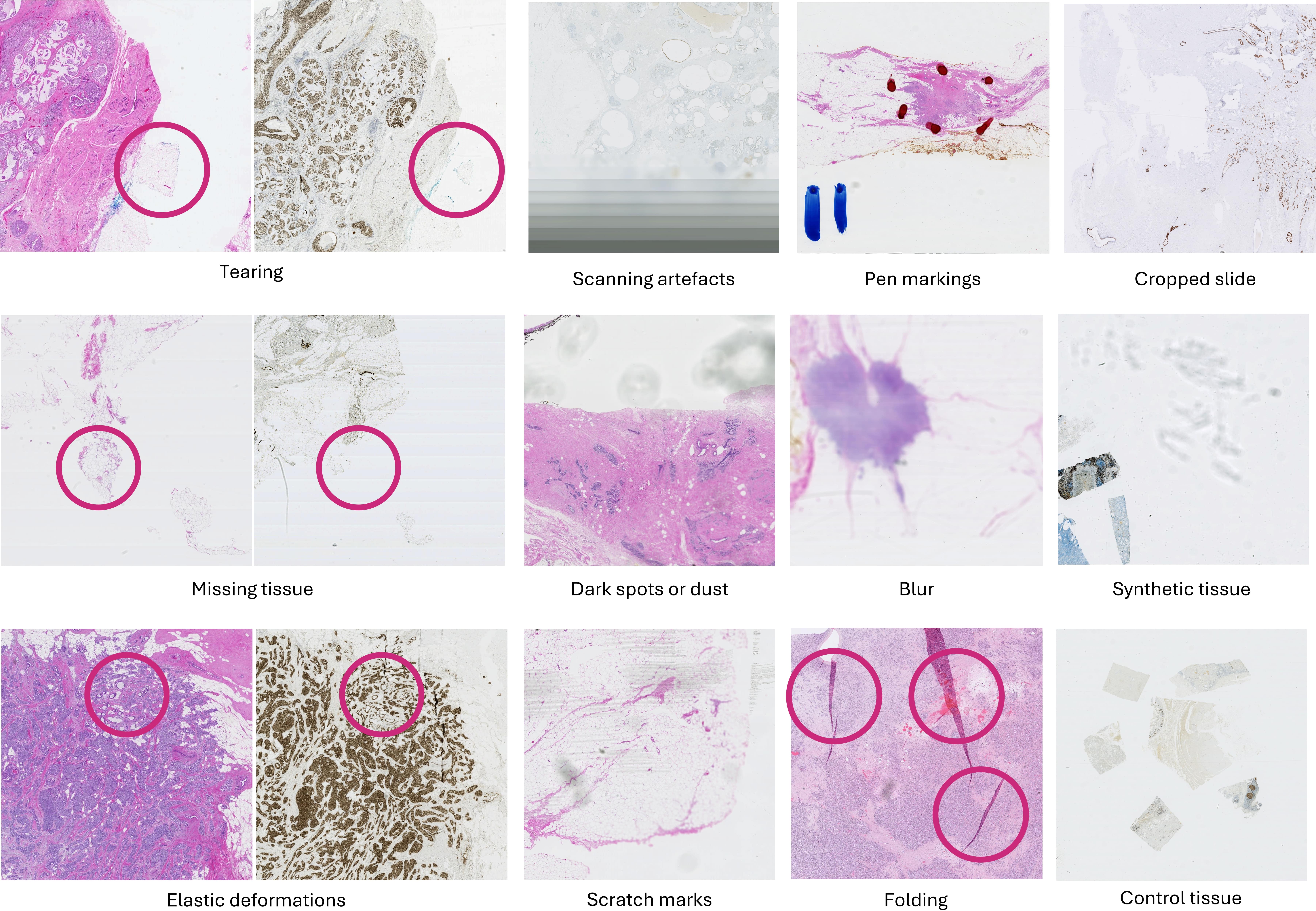}
    \caption{Example images depicting challenges for WSI registration due to imperfections in the digitisation process. These artefacts have been sampled from the ACROBAT 2023 challenge dataset \cite{weitz_acrobat_2023}.\label{fig:ChallengeAll}}
\end{figure*}

The process of slide preparation and digitisation can also introduce multiple challenges that can significantly impact the accuracy and reliability of image analysis. During slide preparation, a tissue section may be placed at an arbitrary orientation compared to other sections of the same tissue block \cite{gurcan2009histopathological}, or tissue contents can change from one section to the next, and there may be unique artefacts on some of the slides \cite{shakhawat2023review}.
For example, elastic deformation, which occurs when the tissue sample stretches or compresses during slide preparation, distorts tissue structures and changes spatial relationships between cellular components \cite{paknezhad2020regional}. Physical imperfections in the slide itself, such as folds that obscure tissue or pen marks, which can be misinterpreted as cellular structures, also challenge the registration of histopathology images. The scanning process can also introduce its own artefacts, like blurriness that obscures details or dust particles that appear as additional structures in the image.  These combined challenges make it difficult to accurately translate the information on the slide into a clean, high-fidelity digital representation. Fig.\ref{fig:ChallengeAll} illustrates a summary of challenges for WSI registration due to imperfections in the slide digitisation process.

Finally, for WSI registration using machine learning algorithms, it is important to have a substantial amount of data with acceptable image quality \cite{faryna2024automatic}. Creating a suitable dataset involves significant expertise and effort, and there are few publicly available datasets for WSI registration. While annotated images are ideal for evaluating models, annotating WSIs with accurate and detailed information is a time-consuming process that often requires input from expert pathologists. Even with expert involvement, establishing a reliable ground truth for model training can be challenging due to potential variations in annotations among different pathologists, which can lead to inconsistent labelling of the same image.

Various approaches have been developed to address the above challenges. For instance, changes in appearance can be handled by colour normalisation techniques \cite{wang2014robust}. 
To address the variations such as deformation during slicing between serial sections in cross-slide registration, one solution is to restain the same section \cite{HyReco}.  However, restaining the tissue sections requires proper training and can be cost-inefficient. Multiplexed fluorescence imaging provides an alternate solution to this problem, but this is relatively new technology, requires staff training, is expensive, and takes considerable time and cost to optimise antibody panels for this technology. 

To address variation in the stains, many researchers employed segmentation approaches \cite{kybic2014automatic, song2013unsupervised}. Additionally, multimodal similarity features, such as structural probability maps \cite{pmid25333109}, and synthesising images with the same domain \cite{roy2023deep, shao2021prosregnet} have also been utilised to mitigate the structural complexity. To address the challenges with the large size of images, two approaches can be adopted: either conducting registration on a different level of downscaled images \cite{wodzinski2024regwsi} or dividing the WSI into smaller patches and registering them individually, followed by the combination of the results \cite{wodzinski2021deephistreg, wodzinski2020unsupervised, roy2023deep}.

\section{Datasets for WSI Registration}

\begin{table*}[ht]
    \centering
\caption{Overview of publicly available WSI Registration Datasets.}
\label{tab:datasets}
\centering  

\begin{tabular}{P{2cm} @{\hspace{0.5em}} p{1.1cm}  @{\hspace{0.5em}} P{4.5cm} @{\hspace{0.5em}} P{5.5cm} @{\hspace{0.5em}} P{2.4cm} P{0.18cm}} \hline
         \textbf{Dataset}&  \textbf{\footnotesize{Size}} &   \textbf{Tissue Type}& \textbf{Staining}&  \textbf{{Scanners}} &  \textbf{Lm} \\ \hline
         ANHIR \cite{borovec2020anhir}&  \footnotesize{480} &   Lung lesions, whole mice lung lobes, mammary glands, mice kidney, colon adenocarcinoma, gastric mucosa and adenocarcinoma, human breast, human kidney&HE, Cc10, proSPC, Ki67, CD31, c-erbB-2/HER-2-neu, ER, PR, cytokeratin, podocin&  Zeiss Axio Imager M1, NanoZoomer 2.0HT, 3DHistec Pannoramic MIDI II, Leica Biosystems Aperio AT2  & Yes\\ 
         ACROBAT \cite{weitz_acrobat_2023}&  \footnotesize{950} &   Breast&HE, ER, PGR, HER2, KI67&  NanoZoomer S360, NanoZoomer XR &Yes \\ 
         HyReCo \cite{van_der_laak_jeroen_hyreco_2021}&  \footnotesize{144} &   N/A&HE, CD8, CD45, Ki67, PHH3 &  N/A &Yes \\ 
         Guiet \& Chiaruttini \cite{TestDatasetRomain}& \footnotesize{2} & Mouse duodenum & Alexa Fluor 555 Azide, primary rat anti BrDU and Harris hematoxylin&N/A & Yes\\ 
        ASHLAR \cite{muhlich2022stitching,ASHLAR_data}& \footnotesize{2} &  Human colon& Hoechst 33342, Pan-CK, $\alpha$-SMA, CD45, CD31, CD3, CD4, CD8a, FOXP3&N/A & No\\ 
         Prostate Fused-MRI-Pathology \cite{madabhushi2016fused} & \footnotesize{28} & Prostate & HE & Aperio & Yes\\
         
         PLISM \cite{ochi2024registered} & \footnotesize{4454 sm, 3417WSIs} & 46 different human tissues & 13 HE conditions  & AT2, GT450, S60, S210, S360, SQ, P, iPhone(6,13), itel, Samsung Galaxy, Moto, Redmi & Yes \\

 \hline
 \multicolumn{6}{P{5.05in}}{\footnotesize{\textbf{Abbreviations:} landmarks available (Lm), smartphone (sm) }} \\ 
 

 
    \end{tabular}
\end{table*}

Challenge competitions have been highly beneficial and have created a positive impact on the advancement of computational pathology. Given its importance, multiple challenge competitions worldwide have focused on WSI registration. These competitions provide large annotated datasets, which makes it easier for researchers to work on high-quality data and propose new methods. This results in the development of state-of-the-art approaches, encouraging further expansion and application. It also helps new researchers in the field to keep track of overall progress through the years and develop new approaches after the competition is over. A list of publicly available datasets has been curated in Table \ref{tab:datasets}.

One of the challenge competitions on WSI registration was ANHIR (Automatic Non-rigid Histological Image Registration) \cite{borovec2020anhir}, hosted at the IEEE International Symposium on Biomedical Imaging in 2019. The competition was specifically organised for histology image registration, drawing 100 registered teams, with 10 actively participating. The results were published for the top 7 well-performing teams. Broadly, most methods employed classic approaches, with the winner achieving a registration accuracy exceeding 98\% for landmarks. An interesting aspect was that the best-performing algorithms were initially developed for different modalities but were fine-tuned for histopathology data. Another noteworthy finding was the minimal difference in both robustness and accuracy among the first six methods, contrasting with significant variations in reported execution speed. The fastest, completed tasks within a few seconds, while the slowest required an hour. None of the evaluated methods took advantage of full-resolution or full-colour information, and there was only one method employing a neural network for registration - a convolutional neural network (CNN) which was the fastest method \cite{borovec2020anhir}. Despite good registration accuracy, with a median landmark localisation error in tens of pixels at the original resolution, even the most effective methods might lack the robustness and accuracy required for routine fully automated use at the nuclei level. However, these methods could be employed in an application with a semi-supervised setting, reducing the effort required for manual refinement.

The ANHIR dataset comprises high-resolution WSIs of various tissue types, including lesions, lung lobes, and mammary glands, at magnifications of up to $40\times$. The original sizes of these images range from $15$K $\times$ $15$K to approximately $50$K $\times$ $50$K pixels. The WSIs were acquired in sets of consecutive tissue slices, with each slice undergoing staining using a different dye, including clara cell 10 protein (Cc10), prosurfactant protein C (proSPC), HE, antigen Ki67, platelet endothelial cell adhesion molecule (PECAM-1, also known as CD31), human epidermal growth factor receptor 2 (c-erbB-2/HER-2-neu), ER, progesterone receptor (PR), cytokeratin, and podocin. A collection of over 50 histological sets is provided, and for user convenience, downscaled versions of the images are included at various resolutions: 100\%, 50\%, 25\%, 10\%, and 5\% of the original size. For evaluation purposes, the landmarks were manually identified in each image, ensuring correspondence within each set.

The ACROBAT (AutomatiC Registration Of Breast cAncer Tissue) challenge \cite{weitz_acrobat_2023, wodzinski2024regwsi}  was organised in 2022 and 2023, running for two editions to date. The challenge was hosted on the Grand Challenge platform \cite{GrandChallenge}, with the ability to view the ranking of the submitted algorithm run through the validation set on the leaderboard. However, in the second edition \cite{wodzinski2024regwsi}, Docker containers were the accepted method of submission via a private upload link. The intended output of registration includes the registered image with the transformed landmarks. The first edition included eight graded submissions \cite{weitz_acrobat_2023}, while the second edition received four submissions \cite{wodzinski2024regwsi}. Arguably, the submitted registration methods represent the cutting-edge advancements of WSI registration, with many participants publishing open-source code and method descriptions.

The ACROBAT dataset is publicly available, and it is based on 4,212 WSIs at $40\times$ magnification from 1,152 breast cancer patients.  HE slides as well as IHC stains including ER, HER2, Ki67, and PR are provided in the dataset, and it is divided into training, validation, and test sets. While the validation set slides were available for participants, the annotations were hidden. Both validation and test sets were manually annotated with landmarks by two experts.

The HyReCo dataset \cite{van_der_laak_jeroen_hyreco_2021}, acquired at the Radboud University Medical Center in the Netherlands, consists of two subsets, A and B. Subset A includes consecutive sections stained with various markers such as HE, CD8, CD45RO, Ki67, and PHH3. These slides were annotated with $11–19$ landmarks per section, totalling $690$ landmarks across all stains. Subset B consists of re-stained slides without corresponding consecutive sections, annotated with an additional $2303$ landmarks. The images in both subsets were digitised at a resolution of $0.24\mu$m/px and are of size approximately $95$K $\times 220$K pixels. To evaluate landmark accuracy, inter-observer and intra-observer errors were measured, indicating high precision. The dataset, along with landmarks, is available under the Creative Commons Attribution-ShareAlike 4.0 International Licence, enabling researchers to utilise it for further analysis and validation.

The Prostate Fused-MRI-Pathology dataset \cite{madabhushi2016fused} is collected for the registration of MRI and histopathology prostate images. It is publicly available through The Cancer Imaging Archive (TCIA) website \footnote{https://www.cancerimagingarchive.net/collection/prostate-fused-mri-pathology/}. This dataset comprises 28 human subjects, each with MR and HE images. Surgically excised prostate specimens were originally sectioned and quartered, resulting in 4 slides per section. Each slide was digitised at $20\times$ magnification using an Aperio slide scanner, producing a set of four WSI images in the *.svs format. These images were digitally stitched together to create pseudo-whole mount sections (.tiff). Cancer presence annotations on the pseudo-whole mount sections were provided by an expert pathologist. Slice correspondences between individual $T2w$ MRI and stitched pseudo-whole mount sections were established and verified for accuracy by expert pathologists and radiologists.

The Pathology Images of Scanners and Mobile phones (PLISM) \cite{ochi2024registered} dataset, recently published, was specifically developed to enhance machine learning methods; variations in the colour and texture of histopathology images undermine the robustness of machine learning models when faced with out-of-domain data. To address this, the PLISM dataset includes precisely aligned image patches from various domains, facilitating accurate evaluation of colour and texture differences.
PLISM comprises images of 46 human tissue types, stained in 13 different ways, and captured using 13 distinct imaging devices. The strength of PLISM is that it includes both WSIs and smartphone images of the same tissue or serial sections of tissue microarray (six smartphones and seven digital scanners). Analysis revealed significant diversity in these variations, particularly between WSI and those from smartphones. A CNN pre-trained on PLISM demonstrated improved handling of domain shifts.

Finally, the datasets ASHLAR (Alignment by
Simultaneous Harmonisation of Layer Adjacency Registration) \cite{muhlich2022stitching,ASHLAR_data} and Guiet \& Chiaruttini \cite{TestDatasetRomain} have limited number of two sample of WSIs.

\section{Evaluation metrics}
Image registration is typically formulated as an optimisation problem or through minimising a loss function. This typically comprises a similarity metric, which compares the aligned images, and a regularisation term that guarantees the smoothness of the transformation. Traditional similarity metrics such as Euclidean Distance or Normalised Cross-Correlation prioritise aligning pixel intensity values or correlations.
However, for WSI, it falls short due to the lack of geometric or physical meaning associated with many common similarity measures. In simple terms, just because two whole slides “look similar” according to a specific measure does not necessarily guarantee they accurately align anatomically.

One reliable approach involves manually identifying a set of corresponding points in both reference and moving images. This technique, known as landmark-based methods, uses these points to assess how well the registration aligns the images. The drawback is that the evaluation of registration depends on the quality of these landmarks. However, the effect of “noisy" annotations can be reduced by marking sufficiently large number of landmark points verified by at least two pathologists. 

For evaluation metrics, Target registration error (TRE) is the most frequent landmark-based metric that quantifies the precision of a registration process. It calculates the misalignment between corresponding landmarks in the transformed and the reference images and is computed as $ \operatorname{TRE}= ( \sum_{1=1}^{N}\left \| p_i-q_i \right \|)/N$, where $N$ is the total number of landmarks, and $ p_i$ and $q_i$ represent the coordinates of the corresponding points in the registered and reference images, respectively.  

Most studies have used TRE in their evaluation process, with different variants of TRE being available for this purpose, \cite{awan2023deep, wodzinski2024regwsi, lotz2019robust}. By normalising the TRE with respect to a predefined length or dimension in the reference image, rTRE calculates the relative error (r$\operatorname{TRE}=\operatorname{TRE}/D$; where $D$ is the reference length or dimension in the reference image used for normalisation). This normalisation allows for a more standardised comparison of registration accuracy across different datasets or applications, particularly when the scale of the images varies. Average rTRE, mean rTRE and max rTRE are other variants of TRE reported for evaluation \cite{borovec2020anhir}.

Apart from TRE, other methods exist for computing distances between landmarks. For instance, Euclidean distance of the position of the landmarks \cite{sorokin2017non}, computing the distance of feature keypoints extracted from automatic methods \cite{faust2022integrating}, the mean absolute error (MAE) of landmarks\cite{shao2021prosregnet} and t-test of the annotations that were manually selected on both images \cite{song2013unsupervised} are also reported. However, these methods are not commonly employed in the literature for histopathology registration.

Since manually selecting keypoints and annotations is labour-intensive and not all the datasets have landmarks, several studies reported the evaluation using similarity metrics. The Dice coefficient and Hausdorff distance are two commonly employed metrics in medical image analysis to assess the similarity or dissimilarity between two sets of points or shapes, particularly in tasks like image segmentation or registration. The Hausdorff distance measures the maximum dissimilarity between two sets of points or shapes by determining the maximum distance between a point in one set and its nearest point in the other set, and vice versa. Conversely, the Dice coefficient quantifies the spatial overlap between two sets of points or regions. Various studies focusing on registration from a segmentation perspective have used the Dice coefficient, Hausdorff distance and mean segmentation offset to report the evaluation results \cite{song2013unsupervised, shao2021prosregnet, lotz2015patch}.

The structural similarity index measure (SSIM) can also be used to evaluate the similarity between two images; it measures the similarity in luminance between corresponding pixels, the similarity in contrast, and the similarity in structure, and is calculated by comparing local patterns of pixel intensities in the two images. Other similarity metrics, including root mean square error (RMSE), peak signal-tonoise ratio (PSNR), mutual information (MI), normalised MI (NMI), normalised gradient field (NGF), normalised correlation correlation (NCC) are also used in several works to quantify the registration precision \cite{roy2023deep, borovec2020anhir}.

The aforementioned metrics are the most frequently used for WSI registration evaluation. However, they do not include all the available methods, and the choice of evaluation metrics may vary depending on how the registration problem is designed. For example, in the case of deep learning models treating registration as a segmentation and classification task, metrics like area under the receiver operating characteristic curve (AUC) or accuracy can also be used for assessment \cite{daly2021convolutional}.

Among all the methods for registration evaluation in histopathology images, landmark-based models are the most effective. This is because they provide precise, localised error measurements. Given the repetitive patterns typical in tissue images, other methods, such as overlay comparison or similarity metrics, may not accurately reflect the quality of the registration.
Although several studies have reported landmark-based error metrics, this method requires accurate identification of landmarks, which can be labour-intensive and is not feasible for large-scale evaluations.

Alternatively, similarity-based evaluations such as MSE, RMSE, and NCC are straightforward to implement and applicable across all methods and datasets. However, they are susceptible to intensity variations and may not be completely ideal for histopathology images. MI, being less sensitive to intensity, proves effective for multimodal image registration. Yet, calculating MI can be computationally expensive, particularly for high-resolution whole images, due to histogram computations and joint entropy estimation involved.

Feature-based evaluation, which compares extracted features like edges and contours from registered and reference images, offers robustness to intensity variations and focuses on significant structures within the images. However, this method heavily relies on the accuracy of feature extraction. Finally, overlap-based metrics, which measure the similarity of segmented regions, are effective and interpretable, especially for the segmentation-based approaches. Nevertheless, they also depend on the accuracy of the segmentation methods.

It is important to note that apart from accuracy, a crucial consideration for a registration method is the running time. A technique that achieves very high accuracy but has a long running time may not be practical for many applications. In contrast, a method with slightly lower accuracy but significantly shorter processing time can be more desirable \cite{mueller2011real}. Due to this trade-off, many studies reported both accuracy and computation time in their evaluations \cite{hoque2022whole, shao2021prosregnet, lotz2015patch}. Borovec et al. \cite{borovec2018benchmarking} published a benchmark paper comparing 11 well-known methods across various aspects. In terms of running time, they reported the running time on the same machine for linear and elastic (free-form) transformations. The minimum average running time for linear transformations is 6.13 seconds, while for free-form deformations, it is 5.92 seconds. The maximum average running time is 787.9 seconds for linear transformations and 17,179 seconds for free-form deformations. Muhlich et al. \cite{muhlich2022stitching} also compared their method to the well-known Microscopy Image Stitching Tool (MIST) \cite{chalfoun2017mist} and found that both methods exhibited similar running times on a single CPU, approximately 300 seconds. In contrast, Song et al. \cite{song2013unsupervised} demonstrated a significant difference in execution time between MI-based and classification-based registration methods. For a specific data group, the maximum execution time was around 2,500 seconds for MI-based methods, whereas it was less than 500 seconds for classification-based methods.

In conclusion, the choice of evaluation method primarily depends on the registration approach employed and the dataset. Each evaluation metric comes with its own set of advantages and disadvantages. Consequently, it is common for studies to report multiple metrics for registration accuracy in addition to computational efficiency to provide a comprehensive assessment. Additionally, given the significance of memory usage and other hardware requirements in registration, including these factors alongside registration results enhances the holistic assessment of a method's performance.

\section{Conventional approaches for WSI registration} 

\begin{figure*}[htbp]
    \centering
    \includegraphics[width=16.5cm]{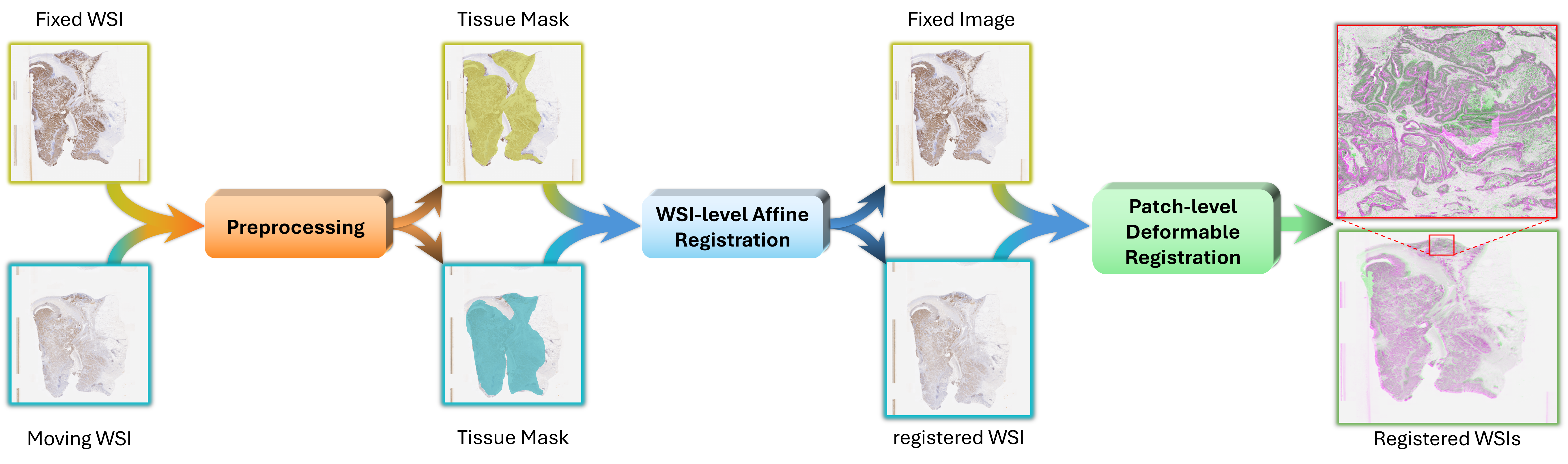}
    \caption{Schematic illustration of a conventional WSI registration pipeline; the first step involves preprocessing the WSI and subsequently, a rough (low pixel resolution) affine transformation is applied to align the reference and moving images approximately. Final step, a non-rigid registration technique is predominantly employed at a high pixel resolution patch level to further refine the transformation and produce the ultimate registration results. \label{fig:pipline}}
\end{figure*}

Based on the level of manual intervention required, there are three main approaches for registration: manual, semi-automatic, and fully automated. Manual registration involves an expert user selecting corresponding points (landmarks) in the reference and moving images for computing transformation parameters. This method usually achieves high accuracy but it is labour-intensive and time-consuming requiring the user to zoom in and out on both reference and moving images and carefully selecting corresponding points  \cite{goshtasby1986piecewise, goshtasby1988image}. Thus, making it impractical for large-scale experiments.

Semi-automatic registration combines manual and automatic processes, with the user providing initial guidance by selecting a few control points, and an algorithm refining the alignment \cite{bergstrom2017robust, bergstrom2014robust}. This reduces user effort, but still requires their input and depends on the quality of the selected points. Fully automated registration is entirely handled by algorithms, eliminating the need for manual point selection but potentially lacking the accuracy of manual methods. In this paper, we focus on fully automated approaches, reviewing and analysing them with a view of their underlying methods.

In general, most of the methods for WSI registration follow a multiresolution i.e., performing registration at multiple resolutions. The main reason is the huge size of images and the inherent pyramid structure which facilitates such an approach. Also, to increase robustness and address hardware limitations, many of these approaches adopt patch-based strategies.
First, they apply pre-processing steps such as intensity normalisation, masking, and segmentation to prepare the images. Next, a coarse but robust alignment, often based on rigid transformations, is performed. Finally, the registration precision is refined further through non-rigid registration. Fig.\ref{fig:pipline} illustrates a conventional WSI registration pipeline.

The automatic image registration techniques published in the literature can be classified based on several factors, including transformation elasticity (rigid, affine, non-rigid), dimensionality of the imaging data (1D, 2D, 3D, ...), imaging modality (unimodal, multimodal), transformation domain (local or global), nature of registration basis (intrinsic or extrinsic), interaction type (interactive, semi-automatic or automatic), parameter estimation (direct or search oriented) and subject (intrasubject, intersubject and atlas) \cite{van1993medical, oliveira2014medical, maintz1998survey}. Broadly, we can classify these into conventional and deep learning-based registration algorithms. The conventional (non-deep learning) methods can be categorised further based on their approaches into four broad groups: intensity-based, feature-based, frequency domain-based and point set-based approaches.

\subsection{Intensity-based registration}
Intensity-based registration techniques align images by comparing pixel intensity values \cite{valsecchi2013evolutionary}. They rely on maximising similarity metrics, such as correlation or MI, between corresponding images. This approach is particularly useful when images exhibit similar distributions but may be subject to differences in contrast, brightness, or noise. The main advantage of these methods is their simplicity and computational efficiency, making them suitable for diverse applications. However, they may struggle with the registration of images that contain significant variations in texture and appearance.

Several similarity metrics have been employed for WSI registration. The Normalised Gradient Fields (NGF) metric is well-known for image registration, particularly when dealing with images that have varying intensity values due to different staining or imaging modalities. The NGF focuses on aligning the gradients of the images rather than their raw intensity values, making it robust to such variations. 
Lotz et al. \cite{lotz2015patch} proposed a novel method for aligning WSI images with six different stains based on NGF. Their method first performs pre-alignment with the affine transformation of the masks computed using the sum of squared distance (SSD) measure. Next, it registers the WSIs at a low resolution with a nonlinear deformation model. It later refines this result on patches by using a second nonlinear registration on each patch. Finally, the deformations computed on all patches are combined by interpolation to form one globally smooth nonlinear deformation. The NGF distance measure is utilised for both nonlinear registrations as the similarity metric.  The authors evaluated their method on 10 WSI pairs of human lung cancer data. The performance of the alignment is measured by comparing manual segmentations from neighbouring slides. This method achieved a significant improvement in alignment accuracy, with at least a 15\% reduction in offset compared to the low-resolution nonlinear registration. The results show that NGF performs effectively for both images with different stains and same-stain registration.

Achieving successful outcomes with NGF, Budelmann et al. \cite{budelmann_histokatfusion_2022} and Liang et al. \cite{liang_improving_2021} also utilised this method to detect similarity for WSI registration.
As part of the ACROBAT 2022 challenge, Budelmann et al. \cite{budelmann_histokatfusion_2022} proposed a Newtonian registration method that relies on optimisation. Firstly, their method segments WSI foregrounds using a CNN and then performs automatic rotation alignment through the centre of mass. Then affine registration is computed using the NGF objective function. A deformable registration utilises the curvature regularisation function in combination with NGF to estimate local deformities at different image resolutions. This model achieved fourth place in ACROBAT 2022.
Liang et al. \cite{liang_improving_2021} also explored aligning nonglobally stained IHC sections through a graph-based registration method by minimising NGF for 3D tissue reconstruction. The dataset comprises five IHC WSIs divided into 315 patches from different tissue types (e.g., prostate, breast, skin, etc.) and 21 IHC stains.

As MI can effectively handle images with varying intensity distributions arising from differences in staining, imaging modalities, or tissue preparation techniques, it has been employed widely in the literature for WSI registration.
Doyle et al. \cite{DOYLE2023100175} proposed a fully automated process for deformable registration of multiplexed digital WSIs. In this work, they generalised the calculation of MI as a registration criterion to an arbitrary number of dimensions, making it well-suited for multiplexed imaging. They also used the self-information of a given immunofluorescence (IF) channel as a criterion to select the optimal channels to use for registration. The results show that their framework registers 6-plex/7-color mIF images with brightfield mIHC images with comparable accuracy. 
With the same idea, Meyer et al. \cite{meyer2006methodology} showed that regardless of whether the moving image was MRI, tissue block image, or stained sample data, maximising MI across all pairs of the image data can register them to the reference HE slides. 
They evaluated their framework on the coronal section of a rat brain containing a 9L Gliosarcoma with an \textit{in vivo} 7T MRI volume of the same brain.

Similar to NGF, MI has applications in both single-modal and multimodal registration tasks.
For instance, in \cite{mosaliganti2006registration} MI was utilised to spatially align HE serial sections, while in \cite{du2005efficient} and \cite{can2008multi} MI was used for aligning HE slides with cryo-sections and fluorescent images respectively. 

Another alternative similarity metric for intensity-based registration is cross-correlation. It measures the similarity between two images by comparing their intensity values at corresponding positions, making it particularly useful for aligning images with similar intensity distributions and structures. 
Kiemen et al. \cite{kiemen_coda_2022} developed a WSI registration method based on cross-correlation similarity for 3D tissue reconstruction. Using seven HE serially-sections images prepared by the authors from pancreatic tissue, a registration algorithm is devised to optimise for NCC. The method employs a batch-based strategy for processing each WSI. Registration results show up to 95\% similarity score between the reference and moving images. Both the dataset and the code are publicly available \footnote{https://github.com/ashleylk/CODA}. In \cite{cooper2007registering}, Cooper et al. proposed to register whole slide multimodal images (HE and PTEN stains) with a sharpness-enhanced NCC metric.
In general, because of its simplicity and computational efficiency, NCC is a well-known metric for registration of WSI in volume reconstruction applications \cite{ma2008automatic, kurien2005three, ruiz2009non, pitiot2006piecewise} where both moving and reference images are the same modality.

Rather than relying on a single similarity metric, many studies used multiple metrics in their approaches. 
Mueller et al. \cite{mueller2011real} applied different similarity metrics for lower and higher resolution for aligning multimodal (HE and IHC) WSI. Their approach first calculates a transformation on lower-resolution images for efficiency. Then, it only applies this transformation to the specific high-resolution area being viewed, allowing for interactive exploration. The key aspect of their work is its ability to balance accuracy and speed. Unlike existing methods, the method prioritises user control by offering a choice between a more precise but slower approach and a faster but less accurate one. The study results show that NCC archives comparable results with MI. In \cite{obando2017multi}, the proposed method involves an initial rigid registration using cross-correlation, followed by a deformable B-spline registration utilising MI. 

In summary, a variety of similarity metrics utilising image intensities have been used for WSI registration. Among the most popular ones, SSD and NCC are well-suited for images with equal or similar pixel intensities, such as those from the same modality. Conversely, MI provides a measure of the degree of concordance between the statistical properties of corresponding regions, which can then be used for registration by optimising the spatial image coordinates. The selection of the appropriate metric should be based on the dataset characteristics and the specific requirements of the WSI registration application. 

In conclusion, intensity-based approaches are generally computationally efficient and straightforward to implement, making them suitable for large datasets. However, they may have limitations. Variations in staining, lighting, and tissue preparation can significantly affect image intensities, leading to registration errors. Additionally, intensity-based methods can struggle with images exhibiting large deformations or non-linear variations, common in histopathology. They may also be sensitive to noise and artefacts present in WSI samples.

\subsection{Feature-based registration}
Feature-based registration operates by detecting and matching distinctive features, such as corners, edges, or blobs, between images \cite{guan2018review}. These features are usually designed to be robust to variations in intensity, contrast, and noise, making them suitable for aligning images with significant differences in appearance. 
Scale-invariant feature transform (SIFT) \cite{lowe2004distinctive} and speeded robust features (SURF) \cite{bay2006surf} are among the most well-known methods for WSI registration.

Saalfeld et al. \cite{saalfeld2012elastic}  proposed to utilise elastic stack alignment (ESA) model. This method initially employs RANSAC \cite{fischler1981random} (Random Sample Consensus;  a robust method for estimating parameters of a mathematical model) to identify a rigid transformation based on SIFT features. Subsequently, a non-linear registration step is performed using normalised correlation and virtual springs to maintain the transformation close to a rigid one. The optimal parameters for the ESA algorithm are then determined through a parallel grid search on a computational cluster.


Theelke et al. \cite{theelke2021iterative} proposed an iterative approach for registering images digitised with multiple scanners. They employed SIFT features to achieve the initial affine transformation, updating the work of Jiang et al. \cite{jiang2019robust}, which only estimated translation. By including rotation and scaling parameters, their method improves the method published by Jiang et al. \cite{jiang2019robust}.

Shafique et al. \cite{shafique2021automatic} proposed an automatic feature-based cross-staining registration method. Initially, image pairs were aligned based on translation, rotation and
scaling and then the registration was performed based on automatic landmark detection in both images using SIFT. Lastly, the fast sample consensus protocol is used for finding point correspondences and aligning the images.

The approach by Paknezhad et al. \cite{paknezhad2020regional} utilises a multiscale attention mechanism for registration, initiating the process over a broad region around the region of interest (ROI) and progressively focusing on a smaller area as resolution increases. This contrasts with existing multiscale algorithms that register the entire tissue at all resolutions. The method's uniqueness lies in its ability to handle significant deformations outside the ROI, enhancing robustness. Additionally, they proposed a method for selecting a subset of SIFT keypoints to avoid non-specific points and contribute to a robust regional registration technique.

Hoque et al. \cite {hoque2022whole} aimed to propose a robust solution for WSI registration using classical techniques. They introduced using gradient SIFT to extract multi-stained features and a new keypoint matching algorithm that integrates scale, keypoint orientation, and position to enhance the accuracy and number of correct keypoint matches.
This framework is the first presented WSI registration which works on large-scale whole slide images without relying on patch-based registration. They compared their results with state-of-the-art methods using SIFT, SURF, and other well-known feature detectors, demonstrating that their model achieves lower running time and better precision and recall.

There are additional features used in the literature that are specific to WSI. For example, Sarkar et al. \cite{sarkar2014robust} proposed using line-based features to align adjacent tissue slides. Their approach involves selecting boundary points around the tissue and employing RANSAC to sample pairs of these points to fit lines randomly. For each sampled pair, a line is fitted, and the sum of gradient magnitudes of points within a ±2 distance from the line is measured. Features such as the angle and distance of the lines are used to compute translation and rotation. After determining the global transformation, normalised correlation based on gradient magnitude images is utilised for more precise local matching in a multi-resolution approach.  In \cite{huang2006fast},  Huang et al. leveraged features like area and eccentricity of segmented red blood cells for registration, and in \cite{cooper2009feature} a broader set of features from segmented HE and CD3 stained images, including centroid, area, eccentricity, and axis orientation were used.

In conclusion, feature-based registration offers advantages in situations where images have intensity variations, making it a popular choice in WSI applications.
Different feature extraction techniques are available, each offering unique advantages in terms of robustness, computational efficiency, and invariance to transformations, thereby enhancing the accuracy and reliability of the registration process. 
While feature-based registration can handle complex deformations and image variability, it may be highly sensitive to feature extraction parameters and the accuracy of the registration heavily depends on the quality and reliability of the detected features \cite{chen2021feature}.
In histopathology images, which often have low texture and repetitive patterns, it may sometimes be extremely difficult to identify meaningful features. Furthermore, a feature selection method that works well for specific staining may not be universally applicable across different types of histopathological images. Additionally, extracting and matching features can be computationally expensive and may require sophisticated algorithms, increasing the complexity and time required for the registration process.

\subsection{Frequency domain-based registration}
Frequency domain-based registration methods convert the images in the frequency domain such as Fourier transform and align the images in the frequency domain \cite{tong2019image}. By transforming images into the frequency space, registration algorithms can exploit similarities or differences in their spectral content to achieve alignment. 
Frequency domain-based registration is particularly effective when images exhibit global distortions or geometric transformations. However, these may struggle with local deformations or nonlinear warping. Moreover, frequency-based methods are sensitive to image artefacts and noise, which can degrade registration accuracy \cite{hoge2003subspace}.
Cross-correlation or phase correlation \cite{kuglin1975phase} methods are commonly used to compute the spatial displacement between images in the frequency domain. 

Muhlich et al. \cite{muhlich2022stitching} introduced a framework utilising phase correlation for the stitching and registration of multiplexed images, aiming to create precise whole slide mosaics. The proposed methodology consists of two stages: initially stitching adjacent tiles within a slide, and registering multiple slides to a reference slide. In both stages, they employed an improved phase correlation algorithm \cite{guizar2008efficient} to align the tiles and slides. Their results show that the framework gained similar performance to existing open-source and commercially available software.

Various techniques, such as Wavelet-Based registration and Fourier-Mellin transform, exist beyond phase correlation but are not commonly used for WSI registration. While frequency-domain methods hold promise for registration, they face limitations for WSI registration that push researchers towards alternative methods. 
The presence of noise and artefacts in histopathology images can significantly degrade the performance because these techniques analyse images in the frequency domain, where noise and artefacts manifest as unwanted frequency components. Noise can introduce high-frequency fluctuations that distort the true frequency content of the image and artefacts can similarly affect the frequency spectrum, leading to inaccurate frequency representations \cite{fan2019brief}. 

\begin{figure*}[h]
    \centering
    \includegraphics[width=17.0cm]{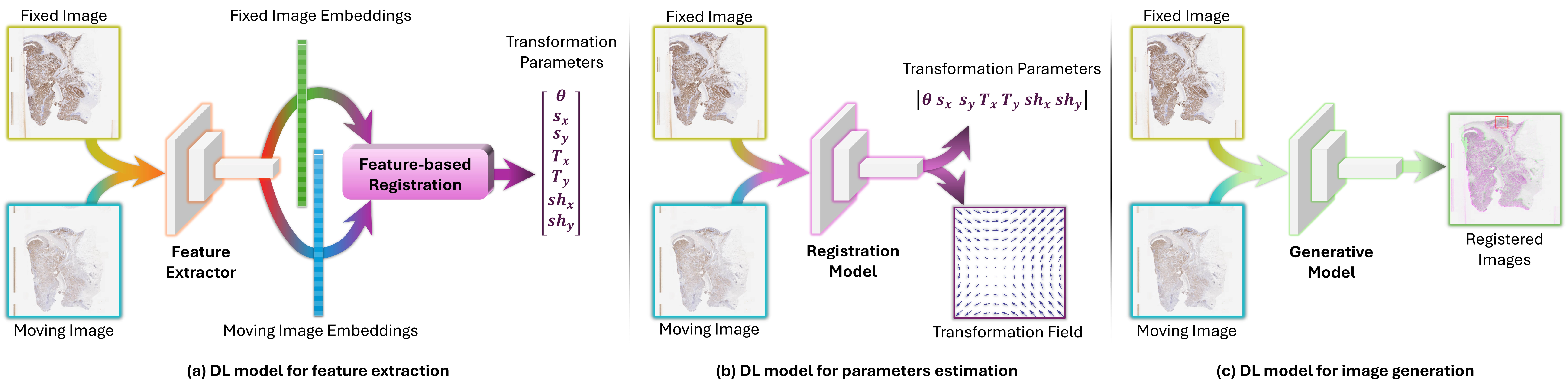}
    \caption{Schematic illustration of a deep learning-based WSI registration model; The models are divided into three different classes based on the network's output. a) a deep learning model that is used as a feature extractor for a feature-based registration approach. The output of the network is a set of feature vectors describing reference and moving images. b) A deep learning model that directly estimates the parameters of transformation between reference and moving images. The output of the network can be either a transformation matrix or a deformation vector field. c) A deep learning model that directly generates the deformed image.   \label{fig:DL_models}}
\end{figure*}

 \subsection{Point-set registration}
 Point set (or cloud) registration approaches compute the spatial transformation that aligns two or more sets of points (clouds). These point set data can be captured directly using specialised cameras or, more commonly, generated through computer vision algorithms including feature extraction techniques such as corner detection.
The process involves an initial alignment to roughly position the point sets, followed by the identification of corresponding points between the sets, often through nearest neighbour search or feature-based matching. Using these correspondences, a transformation is estimated and optimised to minimise the distance between matched points, with methods like Iterative Closest Point (ICP) or Coherent Point Drift (CPD) being commonly employed.

Zhang et al. \cite{zhang2020point} proposed a point set multi-stained image registration method based on SURF features consisting of three steps: extracting matching point, pre-alignment consisting of rigid and an affine transformation on the coarse level, and a B-spline and ICP registration optimised by the extracted points.
Similarly, in \cite{shojaii2009novel} the ICP
algorithm followed by thin-plate splines technique has been used to register the histology sections.
Jeyasangar et al. \cite{jeyasangar2024nuclei} proposed to use the spatial distribution of nuclei for their point set registration approach. 

In general, point set registration methods offer several advantages including their ability to accurately align complex shapes or objects and being relatively straightforward to implement, especially with algorithms like ICP. However, these methods can have disadvantages, including susceptibility to local minima, especially in cases of poor initial alignment or noisy data points. \\

Conventional registration methods, mostly built upon energy optimisation techniques \cite{song2017review}, have long dominated the field of WSI registration, and they are contributing to the vast majority of existing solutions \cite{HOQUE2022105301, lotz2015patch, solorzano2018whole}. However, these methods are not without their trade-offs. On the positive side, their strength lies in their theoretical foundation, providing well-defined optimisation goals and interpretable results. Additionally, their robustness to varying image conditions makes them adaptable to different scenarios. However, their reliance on simplifying assumptions can lead to inaccuracy in complex deformations, and their computational complexity can pose limitations for real-time applications or resource-constrained environments. Furthermore, in some cases, they encounter limitations in automation, requiring manual intervention or fine-tuning. For instance, feature-based approaches necessitate careful selection of feature extraction methods, which can significantly impact registration accuracy \cite{guan2018review}. 

Overall, while conventional methods hold a prominent position in WSI registration, the advancements in deep learning offer greater automation and produce promising results in medical image processing, they are increasingly making their mark in this field, gaining more attention and recognition \cite{wodzinski2024regwsi, wodzinski2021deephistreg}.

\section{Deep learning-based approaches for WSI registration}\label{Sec:deepLearning}

\renewcommand{\arraystretch}{1.4}

\begin{table*}
\caption{Overview of papers using deep learning for WSI registration. \label{tab:alldeepworks}}
\begin{tabular}{P{0.9in}P{1.1in} P{1.0in} P{1.4in} P{1.59in}} 
\hline
Reference &Approach & Model  & Staining & Evaluation metrics \\ \hline

Roy et al. \cite{roy2023deep} &  Parameter estimation, translation  & CNN, CycleGAN & HE, Ki67, PHH3 &  NCC, SSIM,  NMI, HD, Dice coefficient  \\

Xu et al. \cite{xu2020effective} & Translation & CycleGAN & HE, Ki67  & The ratio of positive cells to all cells \\

Daly et al. \cite{daly2021convolutional} & Translation  &  CNN & HE, Nissl & Accuracy, AUPRC, AUROC  \\

Shao et al. (2021a) \cite{shao2021prosregnet} &  Feature extraction, parameter estimation &  CNN, ResNet-101 & whole-mount HE & Dice coefficient, hausdorff distance, TRE \\

Shao et al. (2021b)\cite{shao2021weakly} & Parameter estimation &  VGG16 & whole-mount HE & Dice coefficient, Hausdorff distance, mean TRE \\ 

Shao et al. (2024) \cite{shao2024raphia} & Parameter estimation, translation  &  VGG16, cycle-consistent GAN & whole-mount HE & Dice coefficient, Hausdorff distance, mean TRE \\

Awan et al. (2018) \cite{awan2018deep} &  Feature extractor  & VGG16 & HE, MLH1, MSH2, MSH6, PMS2, CK8/18  & rTRE  \\ 

Awan et al. (2023)\cite{awan2023deep}  & Feature extractor  & CNN &  HE, MLH1, MSH2, MSH6, PMS2, CK8/18 & rTRE  \\

Kondo et al. \cite{kondo_two_2022} & Feature extractor & U-Net, CNN & HE, ER, PR, HER2, KI67 &  rTRE  \\

Santi et al. \cite{santi_nemesis_2022}& Feature extractor & MLP & HE, ER, PR, HER2, KI67  &  rTRE  \\


Zhao et al. \cite{zhao2019unsupervised} &Parameter estimation & CNN & CC10, HE, Ki67, CD31, ER, PR, CKs, Podocin, HER-2-neu & Average, median rTRE  \\

Gatenbee et al. \cite{gatenbee2023virtual} & Feature extractor   & VGG & HE, Brightfield, IHC, IF & Median rTRE   \\

Marzahl et al. \cite{marzahl2021robust}& Feature extractor & GNN & HE, ER, PR, HER2, KI67 & rTRE  \\

Ge et al. \cite{ge_unsupervised_2022} & Feature extractor & CNN &  HE, Cc10, proSPC, Ki67, CD31, HER-2-neu, ER, PR, CKs, podocin & rTRE  \\

Mahapatra et al. \cite{mahapatra2020registration}& Feature extractor & U-Net &  HE, CC10, proSPC, Ki67, CD31, HER-2-neu, ER, PR, CKs, podocin  &  rTRE \\

Wodzinski et al. (2020a) \cite{wodzinski2020unsupervised}&  Parameter estimation & U-Net & SP-C, Ki67, CC10, HER2, PR, ER, CD31,
CKs, HE & rTRE \\

Wodzinski et al. (2020b) \cite{wodzinski2020learning}&  Parameter estimation & ResNet-101, spatial transformer & SP-C, Ki67, CC10, HER2, PR, ER, CD31,
CKs, HE & rTRE \\

Wodzinski et al. (2021) \cite{wodzinski2021deephistreg}&  Parameter estimation & ResNet-101, spatial transformer & SP-C, Ki67, CC10, HER2, PR, ER, CD31,
CKs, HE & Average, median rTRE \\

Wodzinski et al. (2022)\cite{wodzinski_aghsso_nodate} & Feature extractor &Fully-CNN, GNN  & HE, ER, PR, HER2, KI67  & Homography and pose estimation AUC \\

Ekvall et al.\cite{ekvall2024spatial} & Feature extractor &AE & HE, ISS, Visium & TRE, aTRE \\

Deng et al. \cite{deng2021map3d} & Feature extraction & GNN & Hematoxylin   and Lectin & TRE \\

He et al. \cite{he2024registration} & Parameter estimation &  CNN & HE & NCC \\ 

Liu et al. \cite {liu2021histopathology} & Feature extraction & CNN & HE & NCC, NMI , MSE  \\ 

Jeyasangar et al. \cite{jeyasangar2024nuclei} & Feature extraction & CNN & HE, PHH3 & TRE, rTRE \\ 
 \hline
 \multicolumn{5}{P{6.5in}}{\footnotesize{\textbf{Abbreviations:} convolutional neural network (CNN), generative adversarial network (GAN), normalised cross-correlation (NCC), structural similarity index measure (SSIM), normalised mutual information (NMI), Mean Squared Error (MSE), area under the precision-recall curve (AUPRC), area under receiver operating characteristic curve (AUROC;AUC), target registration error (TRE), graph convolutional netowrk (GNN), multi layer perceptron (MLP), autoencoder (AE), in situ sequencing (ISS), accumulated TRE (aTRE)  }}
\end{tabular}

\end{table*}

We classify a registration method as a deep learning-based registration approach that takes the reference and moving images as input to a neural network and generates the output that is used in the registration pipeline. Based on the network output, we classify the approaches into three main categories as explained in Fig.\ref{fig:DL_models}. 
(i) Feature extractors, where the network is used for feature extraction and the output is feature layers of the network. These features are then used to align the images using a conventional image registration model.
(ii) Parameter estimation registration models, where the network output is the parameter of the transformation. These models are usually regression networks that directly compute various parameters based on the type of transformation.
(iii) Generative-based registration, where the output is an image, which can be either the translated image directly or an image used for registration purposes. Such models are mainly generative neural networks.

Although deep learning for WSI registration is a relatively recent development, with limited prior research, the number of studies in this area is growing, and it is gaining increasing attention. Table \ref{tab:alldeepworks} provides a summary of research papers that utilise deep learning models for WSI registration.
In the following section, these research papers have been explored in detail based on the three categories mentioned above.

\subsection{Feature extractor deep models}

\newcolumntype{P}[1]{>{\RaggedRight\footnotesize}p{#1}}

Several studies have employed deep learning models for extracting feature descriptors which are robust to a range of variations in multimodal images. These feature descriptors are then used for medical image registration using classic feature-matching techniques \cite{haskins2020deep, boveiri2020medical}.
Awan et al. \cite{awan2018deep} proposed a convolutional autoencoder to learn the feature representation of the input image pair. The features derived from the encoder of the trained autoencoder were then utilised to identify the optimal transformation through gradient descent. In another study, Awan et al. \cite{awan2023deep} used features extracted by VGG \cite{simonyan2014very} network for the estimation of transformation parameters. The method comprises three primary stages, initial processing, rigid alignment, and non-linear registration. In the processing phase, tissue masks were generated for a pair of images and adjusted to appear more alike using histogram matching. The rigid alignment was estimated using the centre of mass and a series of rotation angles, and refined by their proposed deep feature based (DBFR) method. In DFBR, multiscale CNN features are extracted for an image pair and used to find the matching pairs by considering the feature points with a small feature distance. Following the application of the DFBR method, any minor offset was corrected by a phase correlation method, subsequently followed by an established non-linear registration technique proposed in Lotz et al. \cite{lotz2019robust}.

Kondo et al. \cite{kondo_two_2022} proposed a two-step method of WSI registration as a submission for the ACROBAT 2022 challenge. The first step included greyscale conversion and estimating both translation and rotation. The second step involved VoxelMorph-based non-rigid registration \cite{voxelmorph_Balakrishnan_2019}, where U-Net \cite{ronneberger2015u} is trained on the ACROBAT training set with applied augmentations. 

Ge et al. \cite{ge_unsupervised_2022} developed a multistain registration method that employed an unsupervised structural feature-guided (SFG) CNN, which was robust to both low-resolution rough and high-resolution fine structural features of tissues. The non-rigid network was pre-trained with the synthetic FlyingChair dataset \cite{dosovitskiy2015flownet}. Then the supervised component of the SFG network was trained on landmarks. The method was developed and tested on the ANHIR dataset, and ranked first.

Gatenbee et al. \cite{gatenbee2023virtual} created the virtual alignment of the pathology image series (VALIS) WSI registration library. It can register brightfield, IHC, and IF WSIs. Being open-source, VALIS implements rigid and non-rigid registration and supports 322 image formats at multiple resolutions and can register more than two images at a time. Their proposed method employed multiple techniques in pre-processing, rigid registration and non-rigid registration to optimise performance and reduce registration error. A notable observation is the combined use of both deep learning methods such as VGG and hand-crafted feature descriptors such as BRISK \cite{leutenegger2011brisk}. The method has been benchmarked against the ANHIR dataset, the ACROBAT dataset and the 3D reconstruction dataset by Kartasalo et al. \cite{kartasalo2018comparative}.

Mahapatra et al. \cite{mahapatra2020registration} employed the idea of using segmentation for the registration of histology images; they proposed to employ segmentation maps integration to facilitate non-linear registration through a self-supervised deep learning approach. The segmentation maps were created by employing K-means clustering on concatenated multi-scale feature maps derived from a pre-trained U-Net segmentation model.

Besides CNNs, few other deep learning architectures are commonly used for feature extraction. One example is the utilisation of a multilayer perceptron (MLP), as proposed by Santi et al. \cite{santi_nemesis_2022}, to enhance the registration process conducted by SIFT keypoints and RANSAC estimation. This work achieved 5th place in ACROBAT 2022.  

Most of the reviewed works employed deep networks for feature extraction, but deep learning models have also been proposed for keypoint detection and keypoint matching. 
Liu et al. \cite {liu2021histopathology} proposed a new feature extraction approach to improve their previous model's accuracy. In their new approach, a CNN detects all the nuclei as keypoints and when it is challenging to detect nuclei, they choose texture information and corner points as keypoints instead. Their results showed that the new future detection approach improves the alignment accuracy compared to their previous method \cite{rossetti2017dynamic} where SURF features were explored for registration. Similarly, in \cite{jeyasangar2024nuclei}, the proposed method is a local-level non-rigid registration that uses nuclei-location-based points for aligning multi-stained WSIs. They exploit the spatial distribution of nuclei extracted by deep Hover-Net \cite{graham2019hover} model, which is prominent and consistent across different stains, to establish spatial correspondence.

In 2018, Detone et al. \cite{detone2018superpoint} proposed a fully convolutional network, called SuperPoint, that computes SIFT-like 2D interest point locations and descriptors. Their approach resulted in state-of-the-art homography estimation, outperforming conventional methods such as SIFT. Later in 2020, Sarlin et al. \cite{sarlin2020superglue}  from the same group proposed a graph neural network (GNN) for learning feature matching, to estimate the 3D structure in geometric computer vision tasks. Their network, called SuperGlue, matches two sets of local features by jointly finding correspondences and discarding non-matchable keypoints. Their results show that, compared to traditional heuristics, this technique learns priors over geometric transformations and achieves state-of-the-art performance.

Wodzinski et al. \cite{weitz2023acrobat} used two methods for ACROBAT 2022 registration competition. For keypoint extraction, they proposed a method relying on SuperPoint and SIFT, whereas for feature matching they proposed utilising SuperGlue and RANSAC. Iteratively, affine registration was applied followed by non-rigid multiscale instance optimisation that optimises for local NCC. Their proposed framework achieved 4th place in the competition.

Marzahl et al. \cite{marzahl2021robust} proposed a registration method that relies on tree-based triangulation to perform registration. 
The method employed patch-based high-resolution rigid and non-rigid registration using SuperGlue and attention-based keypoint matching. Their method achieved first place in the competition ACROBAT 2022 challenge. However, their average registration time was five hours per WSI (using NVIDIA GeForce 3080 Mobile, 8GB, 8-core Intel i7-11800H setup). Interestingly,  Deng et al. \cite{deng2021map3d} also proposed a method for 3D whole slide reconstruction using SuperGlue. They tested their model with both SuperGlue and SIFT keypoints and reported that both approaches yielded similar registration errors for their dataset. 

In addition to the standard widely-used neural network architecture such as VGG and ResNet, some researchers have used custom architectures. For instance, in a recent work by Ekvall et al. \cite{ekvall2024spatial}, an autoencoder was utilised for keypoint detection and matching for tissue image registration. The authors employed two separate autoencoders, each trained independently on the reference and moving images. Keypoints were extracted by identifying the latent space of the autoencoders and pinpointing the heatmaps by locating the pixels with the highest intensity for each heatmap. The final registration was performed using thin-plate splines. Their method was evaluated for 3D modelling, multimodal data alignment, and single-modality data registration. The results demonstrated enhanced stability and efficiency of their approach across modalities such as Visium, HE images, and \textit{in situ} sequencing.

\subsection{Transformation parameters estimation}
Most parameter estimation registration methods utilise deep learning as a regression model to estimate the parameters of the transformation \cite{roy2023deep, shao2021prosregnet}. 
Compared to feature extractor deep models, employing CNN to predict transformation parameters for pairs of images with significant deformation, poses greater challenges. This is attributed to the necessity of known correspondences and ground-truth parameters for CNN training, which are often unavailable. The high degree of freedom in potential values of transformation parameters adds to the complexity. One primary solution can be creating a dataset with known deformations \cite{shao2021prosregnet}. This entails synthesising data by applying various transformations to the images in the dataset. With the parameters of the transformations available, it becomes feasible to converge the network in a supervised or unsupervised manner.

Roy et al. \cite{roy2023deep}, employed a four-layer fully convolutional network (FCN) as a regression model to estimate the deformation vector field (DVF) parameters of the transformation in an unsupervised approach. 
To perform multiscale registration, the DVF parameters were extracted from three different convolutional layers of the FCN model. Intending to register HE to IHC staining images, their proposed method performed a pre-alignment through a global affine spatial transformation at a lower image resolution, and then the computed transformation was mapped to the full image resolution level. To further improve accuracy, they performed patch registration followed by a combined registration for patches to achieve whole slide registration. The NCC loss function \cite{briechle2001template} penalised differences in appearance between the reference and moving images. Comparative results with state-of-the-art methods revealed relatively promising outcomes in their study.

Wodzinski et al. \cite{wodzinski2020unsupervised} proposed using an encoder/decoder U-Net-like architecture to generate the displacement field for the transformation of affine registration. One year later, the authors proposed an unsupervised registration framework \cite{wodzinski2021deephistreg} based on their other previous contributions \cite{wodzinski2020learning} but modified and improved it. Their new pipeline consists of pre-processing, initial alignment, affine registration, and finally non-rigid registration. The initial alignment was performed by aligning the reference and moving centroids. Then, the reference was transformed by the translation vector between the centroids, followed by an exhaustive rotation angle search. They adopted a variation of ResNet-101 \cite{he2016deep} for affine parameter estimation. Non-rigid registration was also performed by a pyramid-based iterative deep registration network. Negative normalised cross-correlation was utilised for both stages of registration. In 2024, He et al. \cite{he2024registration} improved Wodzinski's work by modifying several convolutional layers and removing certain modules. These changes reduced the number of parameters, resulting in increased efficiency and robustness.

In a study by Zhao et al. \cite{zhao2019unsupervised}, the authors proposed a novel unsupervised learning approach (volume tweening network; VTN) for 3D medical image registration using a CNN within an end-to-end framework. This method introduces three innovative technical components: (i) an end-to-end cascading scheme that effectively addresses large displacements; (ii) an efficient integration of an affine registration network; and (iii) an additional invertibility loss that promotes backward consistency. Although the method was primarily presented for CT scan and MRI, it was utilised on the multistain histology dataset from the ANHIR challenge, and was the quickest method. Nonetheless, due to its limited generalisability, it did not rank among the top performers regarding registration error, placing 6th out of the 10 participating teams.

In the majority of the reviewed parameter estimation methods, a deep network was trained in an unsupervised manner. In the work by Shao et al. \cite{shao2021prosregnet}, the authors create their ground-truth labels by applying various transformations to their dataset. The model is proposed for multimodal image registration (HE and MRI of prostate cancer) and they trained the network by real and the synthesised transformed image. Their proposed framework consists of a CNN for feature extraction and another CNN for parameter estimation, with these networks operating in separate stages. Following the approach of a geometric image registration study \cite{rocco2017convolutional}, the authors cropped the third layer of ResNet-101 \cite{he2016deep} to extract features from both HE and MRI. These features were then fed into a regression network comprising two convolutional layers and one fully-connected layer, ultimately estimating the final transformation parameters. The registration process was performed in two steps: (i) the estimation of an affine transformation where the regression network outputs six parameters, and (ii) the estimation of a non-rigid transformation using thin-plate spline, where the regression model outputted 72 parameters. Both regression networks were trained in a supervised manner. In the same year, Shao et al. \cite{shao2021weakly} proposed another deep learning model for the same problem, this time using VGG16 in a weakly supervised manner. In their work, the VGG16 model is concatenated with a classifier network composed of dense layers, which generates a vector that parameterises a geometric transformation between the two images.


\subsection{Generative models}
Generative deep learning models have reinforced the process of translating images from one modality to another, and it has found application in various medical image registration, such as transforming CT data into MR data \cite{qin2019unsupervised, wei2019synthesis}.  In the majority of the approaches, generative adversarial networks (GANs) \cite{goodfellow2020generative} are employed to effectively translate images between modalities. This technique is also applicable to WSI, simplifying the registration process by translating from one modality, such as HE, to another modality, like IHC staining. Despite its ease and full automation, this approach requires careful consideration of certain concerns. Most notably, the modalities are often unpaired, since each pathology tissue slide is typically stained only once in clinical practice -unpaired images imply that there is no direct, pixel-to-pixel correspondence between the two images and individual pixels in one image might not have an exact equivalent in the other. The unpaired nature also adds to the challenge of obtaining accurate landmark pair annotations, making it expensive and labour-intensive. To overcome the challenge of unpaired modalities, CycleGAN \cite{zhu2017unpaired} was proposed as a solution, employing two GANs to construct cross-domain mappings. Its objective function introduced an extra term to the conventional GAN loss, ensuring that the generated outputs accurately reflect the reference domain.

Several researchers utilised CycleGAN to cope with the unpair issue of the registration in medical images \cite{tanner2018generative, xu2020adversarial, yang2018unpaired, hiasa2018cross}. As an example, authors in \cite{wolterink2017deep} employed CycleGAN for translation between unpaired CT and MRI. The objective of their work was to convert 2D brain MR image slices into 2D brain CT image slices. Quantitative results showed that the model synthesised CT images closely similar to reference CT images, and the model had superior performance compared to a GAN model trained with paired MR and CT images. 

However, for WSI registration, a few studies have leveraged CycleGAN to facilitate image translation. Xu et al. \cite{xu2020effective} pioneered the use of CycleGAN to generate synthesised images. 
The objective of their study is to analyse histology images for cancer tumour detection. The author highlighted that a notable percentage (about 5–10\%) of patients with tumours cannot be identified using HE-stained images alone, and challenging tumours can be detected exclusively through IHC detection technology. Hence, there is a need to produce IHC staining with Ki67 from HE images to enhance tumour detection capabilities. Their proposed framework included CycleGAN as the base architecture for handling unpaired and unannotated data. In contrast to the conventional CycleGAN approaches, their adversarial training process incorporates class-related information, enhancing performance by aligning class-related feature vectors. Therefore, the inputs of their CycleGAN were Ki67 patch, HE patch and a class-related feature vector extracted from a multiple instance learning classifier. 

Roy et al. \cite{roy2023deep} utilised a modified CycleGAN as a pre-processing step to generate IHC staining from HE images. Regarding their loss function, the authors made adjustments to the traditional CycleGAN; since the cycle consistency loss alone could not ensure feature or structural similarity between synthesised and real images, they added a constraint to ensure identical features in both images. This constraint involved extracting features from a pre-trained VGG16 \cite{simonyan2014very} network and a new term for the loss function; the new term was defined based on the Euclidean distance between feature vectors extracted from real and synthesised images. It is worth noting that the authors added the constraint of identical features directly to the loss function in contrast to the Xu et al. \cite{xu2020effective} work where the features related to the tumour class incorporated into the CycleGAN input without changing the loss function.

Based on the reviewed literature, it is evident that CycleGAN forms the foundation of most image translation-based models. However, it can be combined with other deep learning models to enhance performance. For instance, Shao et al. \cite{shao2021prosregnet} proposed a model for MRI-histology registration, both serving as parameter estimation models that involved several manual preprocessing steps. In their recent work \cite{shao2024raphia}, they aimed to develop a fully automated registration pipeline for MRI-histopathology registration. They used the same custom VGG16 network as before but enhanced it with a custom CycleGAN (geometry-consistent generative adversarial network; GcGAN) model for image-to-image translation to improve their results. A notable aspect of their work is the flexibility it offers users to correct or refine AI predictions at various intermediate steps, such as estimating the rotation angle and horizontal flipping of each histopathology slice, determining MRI-histopathology slice correspondences, and segmenting the prostate on MRI images. Additionally, their code is publicly available, facilitating further exploration and potential improvements \footnote{https://github.com/pimed/RAPHIA?tab=readme-ov-file}.

Addressing common coordinate registration, Daly et al. \cite{daly2021convolutional} introduced GridNet, a deep learning model designed to perform registration based on the features of corresponding regions in both reference and moving images. GridNet comprises two CNNs, one for classification and another for segmentation. Although their proposed approach can achieve high registration accuracy, the complex structure of the network requires careful consideration during training. Furthermore, the input data needs specific sampling and pre-processing using solid-phase capture methods. This involves loading tissue onto slides specially printed with a regularly spaced array of discrete capture areas, each covering a fixed area of tissue.

In summary, deep learning is rapidly gaining attraction for WSI registration, demonstrating its potential to tackle the challenges associated with conventional methods. They can be integrated at various stages of the registration pipeline, offering complementary support to existing techniques or even performing the registration independently. While these methods hold significant promise due to their speed and ability to automate the process, they require substantial datasets with accurate labels for training. Additionally, interpreting the inner workings of deep learning models can be complex. Despite these limitations, the continual advancements in deep learning offer an exciting avenue for achieving robust and efficient WSI registration.

\section{Available software and frameworks}
Recent advancements in computer tools for histological images led to the emergence of software solutions for registration.
ImageJ \cite{schneider2012nih, collins2007imagej}, a leading software in biological image analysis, offers various plugins like StackReg, TurboReg, MultiStackReg, Linear Stack Alignment with SIFT, TrakEM2, and BunwarpJ, each utilising different registration methods.
QuPath \cite{bankhead2017qupath} is another open-source image analysis tool that provides an extension for histological image registration and alignment. Although QuPath is well-known for WSI processing, for registration, it lacks the flexibility of integrating new algorithms using a single platform, necessitating manual registration if the results are unsatisfactory.

TIAToolbox \cite{pocock2022tiatoolbox} offers a comprehensive API for computational pathology tasks, including data loading, preprocessing, inference, post-processing, and visualisation. It facilitates WSI registration through deep feature-based non-linear registration followed by non-rigid alignment using SimpleITK.

WSIReg \cite{patterson_wsireg_2020} a Python library, enables multimodal or mono-modal WSI registration using elastix, employing a graph-based approach for defining transformation paths. It supports linear and non-linear transformation models, as well as the transformation of associated data like masks and shape data. Additionally, it exports registered images in pyramidal OME-TIFF or OME-zarr formats for compatibility with various viewing software, while also providing efficient interpolation and support for reading native WSI formats like .czi and .scn. The main disadvantage of WSIReg is that it does not offer a graphical user interface and requires programming skills to run the application.

In 2022, Chiaruttini et al. \cite{chiaruttini2022open} introduced Warpy, which is a semi-automatic workflow designed for Fiji-based WSI registration and analysis in QuPath. Warpy can handle slides with complex transformations using ImgLib2, elastix, and BigWarp.
ASHLAR \cite{muhlich2022stitching} integrates multiplex images via a 3-step approach: stitching adjacent images in the first cycle, registering subsequent cycle images to the initial one, and producing a multidimensional mosaic image. It supports various microscope image formats and exports OME-TIFF images, all while being implemented in Python.

VALIS \cite{gatenbee2021valis} is another automated pipeline for WSI registration, utilising rigid and/or non-rigid transformations, and compatible with over 300 image formats. Its process includes the conversion of images to numpy arrays, single-channel processing, feature detection, optimal image ordering, rigid registration towards a reference image, and non-rigid registration. Additionally, it estimates errors by calculating the distance between registered matched features in full-resolution images.

Recently, Escobar et al. \cite{escobar2024mmir} introduced MMIR, which is open-source software for the registration of multimodal histological images. MMIR provides a range of visualisation tools, a project manager, and an algorithm manager through a plugin-based architecture. It has a user-friendly graphical interface, which eliminates the need for users to manually modify registration parameters. 
Key features of MMIR are a straightforward visualisation interface that operates independently of external software like QuPath or ImageJ, cloud collaboration capabilities, and automatic transformation of a pyramidal organisation across all images. 

Another software developed by the Fraunhofer Institute group is MEVIS \cite{mevis}, a sophisticated software that provides solutions for image and data-supported early detection, diagnosis, and therapy for medical applications.
Their registration platform is designed to facilitate the alignment of various medical images, such as CT, MRI, PET scans, and histopathology images enabling clinicians and researchers to compare and analyse these images with high precision. MEVIS offers a range of advanced features, including support for both rigid and non-rigid registration techniques, which are essential for accurately aligning anatomical structures that may vary in shape or size between scans. The software is known for its user-friendly interface and robust performance, making it a valuable tool in diagnostic imaging, treatment planning, and research. 

In addition to these tools and software, there are platforms like MATLAB \cite{MATLABImageReg} and Python that offer libraries for image registration, such as SimpleElastix \cite{SimpleElastix} and scikit-image \cite{skimageregistration}, which are primarily designed for medical image registration but can also be applied to WSI. In general, each of these tools and software can excel with certain sets of images while performing poorly with others.

To summarise, the software for WSI registration should be a user-friendly versatile tool.  This means it should seamlessly integrate new methods, like deep learning, to stay at the forefront of the field. Additionally, it should empower users to experiment with different registration approaches to find the best fit for their specific needs. To ensure a smooth workflow, the software should have acceptable execution times, meaning it should not take an unreasonably long time to complete tasks.  Finally, for broad applicability, the software should be able to support a variety of image formats commonly used in histological studies.

\section{Discussion and future directions}

 The registration of WSIs is a challenging task due to the gigapixel scale, differences between the appearances of differently stained tissue, changes in appearance, structure and morphology between tissue regions in non-consecutive sections and the introduction of artefacts, tears and deformations during processing of the micrometre-thin tissue sections. While some of these challenges have been addressed in existing literature, several specific issues remain unresolved and require further research:

\begin{itemize}
    \item Deformable and non-rigid transformations with missing data: Developing a method that can accurately align WSIs despite the complexities introduced by deformable and non-rigid transformations, as well as missing and incomplete data still remains challenging to resolve.

    \item Patch-level vs. whole slide-level registration: Developing a pipeline that provides user control for handling both patch-level and WSI-level registration, while also providing users with control to choose and switch between different levels of registration remains challenging.
    \item Multi-slide registration: When dealing with multiple serial sections of the same tissue specimen, particularly for 3D reconstruction there is a need for a framework capable of handling the multi-slide complexities. 
    \item Artefact handling: Developing a registration method that can effectively handle artefacts during the registration process remains challenging. A robust method must be capable of identifying and compensating for these imperfections to ensure accurate registration.
    \item Evaluation metrics: Developing a universal metric that accurately and fairly evaluates all models while maintaining relevance across varied applications is a complex task.    
\end{itemize}

These challenges persist, and while current studies have made significant efforts to address these, there are some promising future directions for further improvement.

Regarding the development of a robust algorithm for WSI registration, although several studies have attempted to address the issue, most of them are designed for specific data types. This means that these algorithms are often optimised for particular datasets and may not generalise well to different types of WSIs. The variability in tissue structures, staining methods, and imaging conditions including missing data, across datasets, makes it challenging to create a single algorithm that can perform effectively across all scenarios. Consequently, there is no publicly available tool that serves as a general-purpose solution for registration, capable of handling all the challenges and coping with missing data. Missing data can significantly affect the performance of an algorithm, as it disrupts the image content, making it difficult to accurately align corresponding regions.

Regarding patch-level or whole slide-level registration, several methods attempt to combine aspects of both approaches, but most existing works primarily focus on one over the other. Deep model training on WSI or patches comes with its advantages and disadvantages. Training WSI registration on a broader field enables detecting larger deformation and helps with the problem of local maxima. However, constructing a deep learning model capable of analysing WSIs imposes a huge challenge; the analysis of multi-gigabyte images increases the network complexity (since it is correlated with the input dimension) and it is resource (GPU and memory) intensive. Dividing the input data into small patches and downsampling are two approaches that can address the complexity issue. Since down-sampling can result in information loss and limit registration accuracy, evidently, most of the reviewed works applied image patches instead of WSI to their network. Qaiser et al. \cite{qaiser2018her} showed that when patch-level labels are accessible, employing patch analysis can train deep learning models to achieve or surpass the accuracy of pathologists. Moreover, WSI-level training can suffer if the dataset is small, in patch-based training each WSI can be divided into many patches, and the model can be trained effectively on a small dataset. Additionally, patch-based training typically has superior local performance compared to WSI training. 

Despite advantages, patch-based registration can suffer from challenges during the patch fusion process; grid-like artefacts can be produced along the patch edges. Although including a large patch overlap before fusion can address the issue, it increases the complexity of the model.
Combining both approaches as a multiscale strategy for registration has been shown to improve performance, as demonstrated in recent works \cite{daly2021convolutional, awan2023deep}. For future work, developing a deep learning-based tool that allows users to choose between patch-based or whole slide-level registration at different stages could be a valuable option. This would enable more accurate results by leveraging the strengths of both methods while maintaining efficiency.

Regarding multi-slide registration, several studies have addressed the challenges associated with aligning multiple slides \cite{muhlich2022stitching, deng2022dense}, but limitations persist. One major challenge is aligning images from different microscopes, which often involves accounting for variations in camera angles caused by the rotation of individual cameras and microscope stages. Additionally, scaling differences can arise even when the same objective is used, due to variations in transfer optics and sensor configurations across different instruments. Developing a model that can handle registration across data from different microscopes is a promising direction for future research.
It is worth noting that with the advent of multiplex imaging, the need for multi-slide registration is decreasing, as multiplex imaging does not require serial sections. However, multi-slide registration remains crucial for accurate 3D reconstruction of tissue structures.

Another promising future direction is the development of methods that can effectively handle artefacts during registration. Some studies have proposed techniques to detect and segment artefacts like folds, ink, bubbles, dust, and pen marks \cite{smit2021quality, hossain2023tissue}, and they are mostly deep learning models. Integrating these capabilities into a registration framework that can automatically perform both artefact detection and image registration would be a significant and valuable advancement in the field.

Regarding the evaluation metrics, diverse measurements have been introduced in various works, depending on the specific architecture and application of the model. Across all models, the landmark-based error consistently serves as a standard evaluation metric.
However, evaluating registration requires a combination of metrics to ensure a comprehensive assessment of alignment quality. The choice of metrics should be based on the specific application and validated across diverse datasets to ensure robustness and generalisation.
Multi-scale quantitative evaluation from patch-level to whole-slide level combined with qualitative assessments, such as visual inspection ensures a comprehensive evaluation. Therefore, it would be beneficial to develop specific metrics to assess the robustness of algorithms in registration as well as in handling common artefacts and staining variations.

The challenges discussed represent the main open areas and their potential future directions. However, there are also additional recommendations that could be valuable for future work.

First, whether the approach relies on deep learning or not, pre-processing plays a crucial role in WSI registration. Tasks such as identifying foreground and background, normalising image intensities, removing artefacts, segmentation and masking, histogram normalisation, rescaling, greyscale conversion, and numerous other pre-processing methods are deployed. The choice of these methods depends on the specific characteristics of the input image and the registration model being utilised. Improving these pre-processing steps can significantly impact the accuracy and quality of the registration results. Similarly, pre-alignment can be a very important stage in WSI registration, and the majority of models incorporate some form of pre-alignment strategy. Common approaches include applying global affine transformations on low-resolution images, utilising phase correlation, employing centre-of-mass alignment, and using feature points. These methods are widely employed in various cases to ensure effective pre-alignment before the registration process. Developing new pre-alignment techniques could be valuable for future work and improving existing models.

Secondly, there are various public datasets for WSI registration, but the majority of studies reported their results for non-public datasets. The datasets utilised mostly had a substantial amount of data, and deep learning models mainly trained on image patches, ensuring a large training dataset. Although it has been demonstrated that networks trained on one dataset can be applied to unseen datasets where the two image domains closely resemble each other \cite{daly2021convolutional}, as most WSI registration approaches require registration with IHC images from various markers, the assumption of close resemblance is not fulfilled in many cases.
Therefore, when recommending a benchmark dataset, it would be beneficial to assemble a diverse collection that includes a wide range of tissue types and staining methods, ensuring generalisation across various pathological conditions. The dataset would be more valuable if it featured high-resolution, gigapixel images that support multi-scale analysis from patch-level to whole-slide level. 
To facilitate robust algorithm development, it is better for the dataset to include accurate expert annotations, ideally with annotated landmarks or keypoints, and to incorporate realistic challenges such as artefacts and non-rigid deformations. Public availability and accessibility of the dataset are also important, along with detailed documentation on image acquisition, staining protocols, and annotation procedures. Encouraging community involvement through collaborative dataset expansion and hosting challenges can further enhance the dataset's value.

Finally, based on the analysis of the literature reviewed in this study, it is clear that deep learning has gained a significant rise in its application for WSI registration, particularly evident in the number of papers published post-2019 and the challenge competitions \cite{borovec2020anhir, weitz_acrobat_2023}. Various neural network architectures have been explored for this purpose, with CNN as the most widely employed for diverse registration purposes. Additionally, generative networks, including GAN and its variants, have been employed for image translation tasks across many studies. Alternative architectures like MLP, U-Net, superGlue, and superPoint find applications in different stages of the registration process. Regarding the training strategies, many of the methods have used pre-trained networks, often supplemented by data augmentation techniques to enhance learning outcomes. \\
However, evaluating the performance of deep learning-based models reveals that they have not yet surpassed substantially over conventional deformable image registration methods. For instance, the deep learning-based approach presented by Awan et al. \cite{awan2023deep} shows performance closely aligned with the state-of-the-art conventional method proposed by Lotz et al. \cite{lotz2019robust} in terms of TRE. Nevertheless, DL-based methods exhibit a significantly faster processing speed, typically by an order of magnitude, attributed to their non-iterative nature and the utilisation of powerful GPUs. Shao et al. \cite{shao2021prosregnet} showed that their proposed deep learning framework can perform registration at least 20 times faster compared to a traditional state-of-the-art algorithm. Consequently, there are numerous opportunities for developing novel deep learning models that could surpass traditional methods and further enhance performance.

\section{Conclusions}
At the time of writing this review, there is a noticeable absence of a comprehensive review that specifically examines WSI registration and explores the use of deep learning approaches in this context. This study addresses this gap through an in-depth review of WSI registration literature, with a majority published within the last 5 years. \\
In conclusion, WSI registration faces different challenges in contrast to traditional medical image registration. The substantial size of WSIs, coupled with the complexities of the nature of the data, adds to the difficulty of achieving a successful registration method. The adoption of deep learning techniques has recently gained interest in WSI registration. While deep learning offers distinct advantages, it also presents its own set of disadvantages and challenges that need to be carefully considered. In the realm of WSI registration, deep learning algorithms are still in their infancy and have not yet surpassed conventional registration approaches mainly due to the sheer size of WSIs. However, its adoption is steadily growing, suggesting potential advancements in the future.

\section*{{Acknowledgements}}
BE is a postdoc on a grant funded by the MRC (MR/X011585/1). SEAR report financial support provided by the MRC (MR/X011585/1) and the BigPicture project, funded by the European Commission. MJ reports financial support provided by the BigPicture project which is funded by the European Commission. NR and FM report financial support from GlaxoSmithKline, United Kingdom, outside of the submitted work. FM is supported by EPSRC grant EP/W02909X/1 and holds a minority shareholding in Histofy Ltd. NR is CEO and CSO of Histofy Ltd.

{
\small 
\bibliographystyle{unsrt}
\bibliography{refs}
}

\end{document}